\documentclass[aps,twocolumn,showpacs,floatfix,groupedaddress,amsmath,amssymb,prx]{revtex4-1}
\usepackage{graphicx}
\usepackage{multirow}
\usepackage{color}
\usepackage{xfrac}

\newcommand{\proot}[2]{\sqrt[\leftroot{-2}\uproot{2}#1]{#2}}
\newcommand{\ket} [1] {| #1 \rangle}
\newcommand{\bra} [1] {\langle #1 |}
\newcommand{\braket}[2]{\langle #1 | #2 \rangle}

\newcommand{\smallmat}[4]{\bigl(\begin{smallmatrix}#1&#2\\#3&#4\end{smallmatrix}\bigr)}

\newcommand{\mypsi}{\ket{\Psi^{\tiny \mbox{bound}}}}
\newcommand{\myphi}{\ket{\Psi^{\tiny \mbox{bulk}}}}

\newcommand{\myphic}{\bra{\Psi^{\tiny \mbox{bulk}}}}

\newcommand{\mypsisym}{\ket{\Phi^{\tiny \mbox{bound}}}}
\newcommand{\myphisym}{\ket{\Phi^{\tiny \mbox{bulk}}}}
\newcommand{\myphisymtopo}{\ket{\Phi^{\tiny \mbox{bulk}}_{\tiny \mbox{topo}}}}

\newcommand{\utopo}{\hat{u}^{\tiny \mbox{topo}}}
\newcommand{\wtopo}{\hat{w}^{\tiny \mbox{topo}}}

\newcommand{\ughz}{\hat{u}^{\tiny \mbox{GHZ}}}
\newcommand{\wghz}{\hat{w}^{\tiny \mbox{GHZ}}}

\newcommand{\zz}{^{\scriptscriptstyle [Z_2]}}
\newcommand{\eref}[1]{Eq.~(\ref{#1})}

\newcommand{\fref}[1]{Fig.~\ref{#1}}

\newcommand{\myg} {\mathcal{G}}

\newcommand{\tr}{\mathop{\mathrm{tr}}}

\newcommand{\ba}{\begin{align}}
\newcommand{\ea}{\end{align}}
\newcommand{\bea}{\begin{eqnarray}}
\newcommand{\eea}{\end{eqnarray}}

\def\id{I}
\def\1{\mat{\id}}
\def\mat#1{\mathbf{#1}}

\newcommand{\sout}[1]{}

\usepackage{array}

\begin{document}
\title{Holographic spin networks from tensor network states}
\author{Sukhwinder Singh$^{1}$, Nathan A. McMahon$^{2,3}$, and Gavin K. Brennen$^{3}$}
\email{Emails: Sukhbinder.Singh@oeaw.ac.at, nathan.mcmahon@uqconnect.edu.au, gavin.brennen@mq.edu.au}
\affiliation{$^{1}$Institute for Quantum Information \& Quantum Optics, Austrian Academy of Sciences, Vienna, Austria}
\affiliation{$^{2}$Center for Engineered Quantum Systems, School of Mathematics \& Physics, The University of Queensland, St Lucia, Queensland 4072, Australia}
\affiliation{$^{3}$Center for Engineered Quantum Systems, Dept. of Physics \& Astronomy, Macquarie University, 2109 NSW, Australia}

%, and (ii) the spectrum of a reduced density matrix in the bulk, which is obtained by tracing out the gauge degrees of freedom, exhibits degeneracies, possibly suggesting an emergent symmetry in the non-gauge sector in the bulk. 
\begin{abstract}
In the holographic correspondence of quantum gravity, a global onsite symmetry at the boundary generally translates to a local gauge symmetry in the bulk. We describe one way how the global boundary onsite symmetries can be gauged within the formalism of the multi-scale renormalization ansatz (MERA), in light of the ongoing discussion between tensor networks and holography. We describe how to ``lift'' the MERA representation of the ground state of a generic one dimensional (1D) local Hamiltonian, which has a global onsite symmetry, to a dual quantum state of a 2D ``bulk'' lattice on which the symmetry appears gauged. The 2D bulk state decomposes in terms of spin network states, which label a basis in the gauge-invariant sector of the bulk lattice. This decomposition is instrumental to obtain expectation values of gauge-invariant observables in the bulk, and also reveals that the bulk state is generally entangled between the gauge and the remaining (``gravitational'') bulk degrees of freedom that are not fixed by the symmetry. We present numerical results for ground states of several 1D critical spin chains to illustrate that the bulk entanglement potentially depends on the central charge of the underlying conformal field theory. We also discuss the possibility of emergent topological order in the bulk using a simple example, and also of emergent symmetries in the non-gauge (``gravitational'') sector in the bulk. More broadly, our holographic model translates the MERA, a tensor network state, to a superposition of spin network states, as they appear in lattice gauge theories in one higher dimension.
\end{abstract}

\maketitle
\tableofcontents

\section{Introduction}
The holographic principle, an anticipated feature of quantum gravity, asserts that at least certain theories of gravity can be described as quantum field theories that live in one less spacetime dimension. For example, in the AdS/CFT correspondence---a concrete realization of the holographic principle---the gravity system lives in a $d+1$ dimensional anti-deSitter (AdS) spacetime and is equivalent to a conformal field theory (CFT) that lives on the $d$ dimensional boundary of the spacetime \cite{AdSCFT, AdSCFTdictionary}. The extra dimension in the bulk spacetime is identified with the length scale of the boundary system, and the renormalization group equations essentially generalize the equations that describe gravity. 

Recently, it has been proposed that the \textit{multi-scale entanglement renormalization ansatz} (MERA) \cite{MERA}---an efficient representation of ground states of local Hamiltonians on a lattice \cite{localHam}---realizes at least some features of the AdS/CFT correspondence \cite{Swingle,MERAHolo,localScaleMERA}.
For example, the MERA representation of the ground state of a one dimensional (1D) quantum lattice system is a two dimensional (2D) hyperbolic tensor network, which also describes the RG flow of the ground state. Specifically, the MERA is based on a real space RG transformation, known as \textit{entanglement renormalization}, that removes local entanglement before coarse-graining the state \cite{ER}. In particular, the extra dimension of the tensor network corresponds to length scale of the 1D system. 

In Ref.~\onlinecite{TNC} one of us introduced a toy model for holography based on the MERA representation of ground states of 1D local Hamiltonians. The model, dubbed \textit{tensor network state correspondence}, illustrates a possible way in which the MERA could encode a dual 2D bulk description of a 1D ground state, however, without paying attention to the presence of onsite symmeties in the boundary theory. In this paper, we generalize the model in such a way that an onsite symmetry at the boundary is gauged in the bulk. In particular, this generalization allows us to establish a connection between the MERA and spin networks as they appear in lattice gauge theories in one higher dimension, while also realizing another important feature of the AdS/CFT correspondence using the MERA.
%A tensor network ansatz for the dual 2D bulk state is constructed by inserting 3-index tensors (with open indices) on the bonds of the MERA tensor network.

\subsection{Tensor network state correspondence}
The basic idea behind \textit{tensor network state correspondence} is that a tensor network (with open indices) can be viewed as a representation of two different quantum many-body states (belonging to two different Hilbert spaces) depending on how a many-body Hilbert space is associated with the tensor network. We refer the reader for details to Ref.~\onlinecite{TNC}. Below, we only briefly summarize the main idea.

The open indices of the tensor network can be associated to sites of a quantum many-body system, specifically, we can use each open index to label an orthonormal basis on a different site of the many-body system. Subsequently, the tensor network defines a quantum many-body state of the system such that the probability amplitude of a given configuration of the sites is obtained by fixing the value of the open indices to correspond to that configuration, and contracting all the tensor networks together by summing over the bond indices. The quantum many-body state obtained following this prescription is referred to as a \textit{tensor network state}. Examples of tensor network states include the MERA, matrix product states (MPS) \cite{Fannes92}, and projected entangled pair states \cite{Verstraete04} (PEPS).

Alternatively, one can associate both the open and bond indices of the tensor network with sites of a larger quantum many-body system, namely, by using each index (open or bond) in the tensor network to label an orthonormal basis on a different site of the system. Subsequently, the tensor network defines a different quantum many-body state whose amplitudes are obtained by fixing the value of \textit{all} the indices of the tensor network and multiplying together the resulting tensor coefficients, one selected from each tensor. We refer to the many-body state obtained from the tensor network in this way as a \textit{tensor network bond state}.

Thus, a generic tensor network (with open indices) can be viewed representing either as a tensor network state or as a tensor network bond state. In Ref.~\onlinecite{TNC} we illustrated that the properties of these two states, which are obtained from the same tensor network, are related together in a systematic way. Thus, a tensor network may be viewed as a `correspondence' between these two quantum many-body states. 

Note that a tensor network bond state may be regarded as a regular tensor network state (where degrees of freedom are associated only with open indices and bond indices are summed over) by modifying the tensor network in a particular way. Namely, by inserting a three index \textit{copy tensor} on each bond of the tensor network, see Ref.~\onlinecite{TNC}.

In Ref.~\onlinecite{TNC}, by applying this tensor network state correspondence to the MERA we obtained a toy model for holography (without considering symmetries). The tensor network state and the tensor network bond state obtained from a MERA correspond to the boundary and dual bulk state respectively. The bulk states obtained from the MERA in this way exhibit some interesting features. First, the bulk states satisfy an area law entanglement scaling \cite{AreaLaw}. Second, the bulk entanglement and correlations are organized according to holographic screens. And third, given the MERA representation of a critical boundary state, the boundary correlators of scaling operators (of the underlying CFT) can be obtained from the expectation value of extended bulk operators in certain dual bulk states.
Some of these results caricature certain features of the AdS/CFT correspondence as described in Ref.~\onlinecite{TNC}.

In this paper, we further develop the toy model. We present a generalized construction of bulk states that retains the three features listed above, but also exhibits new features that result from the presence of a global onsite symmetry in the boundary description.

\subsection{This paper: generalized holographic correspondence in the presence of onsite symmetries}

In the AdS/CFT correspondence, a global onsite symmetry of the boundary system generally translates to a local gauge symmetry in the dual bulk description \cite{AdSCFTdictionary}. Consequently, the bulk description generally consists, in addition to gravitational and matter degrees of freedom, gauge fields that are described by the boundary global symmetry group. In the quantum gravity regime, the bulk state is expected to be entangled between all these degrees of freedom. In this paper, we describe how these features of the AdS/CFT correspondence can be realized within the framework introduced in Ref.~\onlinecite{TNC}. 

Symmetries must be properly accounted for in the RG description of a quantum many-body system, in order to reproduce the large-scale properties effectively. For example, consider 1D local, gapped Hamiltonians that have a global onsite $Z_2 \times Z_2$ symmetry corresponding to $\pi$ rotations about two orthogonal axes. These Hamiltonians can be partitioned into two different equivalence classes or quantum phases, each with distinct large length scale properties: the Haldane phase and the trivial phase \cite{AKLT,SPTP}. %More specifically, a Hamiltonian in either phase cannot be adiabatically connected to the other phase by local unitary transformations that commute with the symmetry. 
More specifically, ground states belonging to the Haldane phase cannot be disentangled to a product state along the RG flow, as long as the RG (entanglement renormalization) transformations protect the symmetry. (Since product states are representative of the trivial phase.) One way to ensure this is to impose that the tensors that implement entanglement renormalization commute with the symmetry. The resulting \textit{symmetry-protected entanglement renormalization} generates a MERA representation that captures both the expected RG flow of the ground state, and also its global symmetry exactly \cite{Singh131}.

In this paper, we consider a local 1D Hamiltonian that has a global onsite symmetry $\mathcal{G}$ (which is not broken in the ground state). We represent its ground state by a \textit{symmetry-protected} MERA and obtain a `dual' 2D bulk state, by extending the construction of  Ref.~\onlinecite{TNC}. The bulk state is decribed by a 2D tensor network that is obtained by inserting a 4-index, \textit{symmetric} copy tensor on every bond of the MERA. Each copy tensor has two open indices, which correspond to bulk degrees of freedom that carry `left' and `right' gauge transformations. Thus, our construction, which takes into account the boundary symmetry $\mathcal{G}$, leads to a bulk state in which the symmetry is gauged, thus realizing the holographic translation of a boundary global symmetry to a local gauge symmetry in the bulk. One may, in retrospect, view the (bulk) gauging of boundary symmetries as an underlying motivation for associating the dual bulk degrees of freedom with the bonds of the tensor network, as opposed to e.g. associating them with the tensors as in previous bulk descriptions of the MERA presented in Refs.~\onlinecite{ExactHolo,HoloCode,HoloRandom}.

\subsection{Connection to lattice gauge theories and spin networks}
More broadly, in this paper, we establish a connection between the MERA and lattice gauge theories (on a hyperbolic lattice) in one higher dimension. In a lattice gauge theory, the degrees of freedom are placed on the edges of the lattice, and elementary gauge transformations act on the sites located immediately around a vertex. In our bulk construction, the bulk lattice overlays the MERA tensor network, after it is embedded in a manifold, and the dual bulk degrees of freedom live on the edges of the bulk lattice (that is, the bonds of the tensor network). Elementary gauge transformations act on the bulk sites located immediately around a tensor. In particular, the two open indices of a copy tensor carry the `left' and `right' gauge transformations respectively. Mimicking this basic setup of a lattice gauge theory allows us to manifest a bulk gauge symmetry, which is seen to be dual to the global symmetry at the boundary.

We show how the bulk states decompose as a superposition of spin network states, as they appear in a 2D lattice gauge theory with gauge group $\mathcal{G}$, where they span the gauge invariant subspace of the Hilbert space \cite{BaezSpinNetwork}. The spin network decomposition allows us to explore further parallels with holography. One, it reveals entanglement and correlations between the gauge degrees of freedom and the remaining bulk degrees of freedom that are not constrained by the symmetry. And second, by exposing the gauge degrees of freedom in the bulk, the spin network decomposition also allows one to calculate expectation values of gauge-invariant observables in the bulk. We also construct a local, gauge-invariant parent Hamiltonian for the MERA bulk states, see Appendix \ref{app:parentHam}.

\subsection{Differences from previous work}
%Before presenting the framework we wish to highlight how our holographic correspondence differs from other related approaches.

A local symmetry also manifests simply in the bulk of a symmetry-protected MERA tensor network representation of a (1D) quantum many-body state with a global onsite symmetry, without reference to a bulk state \cite{Singh131,Singh,SinghSU2}. However, we emphasize that in this paper we implement the holographic gauging of a global boundary symmetry \textit{more manifestly} by means of boundary and bulk quantum states, while Ref.~\onlinecite{Singh131} describes the bulk gauging of the boundary symmetry only at the level of the tensor network. Having access to a bulk quantum state, in which the boundary symmetry appears gauged, allows us to probe interesting features in the bulk to explore further connections with holography. For example, we explore the entanglement and correlations between the gauge and the non-gauge degrees of freedom in the bulk, and emergent symmetries in the non-gauge sector of the bulk. On the other hand, no such notions can be defined when the symmetry is gauged only at the level of the tensor network.

In our construction, the local gauge symmetry is hardwired into the bulk tensor network ansatz. Explicit tensor network representations of quantum many-body states with a local gauge symmetry have been presented by other authors \cite{Banuls2013,LucaGauge,Rico2014,Buyens2014,GaugePEPS,Zohar2015,BiancaTN}. The 2D bulk states that we construct here indeed belong to the gauge-invariant tensor network ansatz e.g. presented in \cite{LucaGauge}. However, in this paper we focus on the construction of a gauge-invariant bulk state from the MERA representation of a 1D ground state, instead of, say, variationally minimizing the energy of a 2D gauge-invariant bulk Hamiltonian.

Ostensibly, our lifting procedure appears similar to the prescription to gauge quantum states presented in Ref \cite{GaugePEPS}. There the authors describe how to gauge the global symmetry of a tensor network state. Specifically, they consider a 2D quantum many-body state represented by a PEPS tensor network and translate it to another 2D quantum many-body state with a gauged symmetry. In contrast, our construction produces a quantum many-body state in one higher dimension. Moreover, our approach is aimed at building a higher dimensional bulk description of symmetric ground states, whereas Ref \cite{GaugePEPS} is not concerned with applications related to holography. 

%Our approach also differs from bulk/boundary type correspondences appearing in condensed matter physics. These are of two types. First is, for example, the appearance of gapless modes on the edge of a topological insulator material \cite{topoInsulator}. Here the boundary mode simply corresponds to excited bulk mode where the excitations above the bulk ground state are localized at the boundary of the material. In the second type of bulk/boundary type correspondence, a boundary Hamiltonian is defined as the logarithm of the entanglement spectrum \cite{} for a virtual bipartition of the ground state, or a boundary state is obtained by tracing out the bulk degrees of freedom.  However, in our description the boundary degrees of freedom are not included in the bulk description and we obtain a pure bulk state from the MERA representation of a pure boundary state. Furthermore, the boundary state is recovered by a global projection on the bulk state and not by tracing out some degrees of freedom. 

Some of the previous proposals for drawing a bulk description from the MERA, those presented in Refs.~\onlinecite{ExactHolo,HoloCode,HoloRandom}, associate the bulk degrees of freedom with the tensors of the MERA. In contrast, here we present a bulk description of the MERA by associating bulk degrees of freedom to the bonds of the tensor network, which is closer in spirit to the organization of the degrees of freedom in lattice gauge theories and allows for a more natural introduction of gauge transformations in the bulk and gauging of boundary onsite symmetries, which appears as a general rule of thumb in the AdS/CFT correspondence.

\subsection{Organization of the paper}
The paper is organized as follows. In Sec.~\ref{sec:boundary} we briefly review the symmetry-protected MERA representation of a 1D ground state that has an onsite global symmetry. In Sec.~\ref{sec:bulk} we describe how to lift the MERA representation to a 2D dual bulk state. In Sec.~\ref{sec:spinnetwork} we describe how the bulk states decompose as a superposition of spin network states. In Sec.~\ref{sec:bulkentanglement}, we present numerical results pertaining to the entanglement and correlations in bulk states dual to the ground states of several critical spin chains of interest. For example, we find evidence for a dependence of bulk entanglement on the central charge of the boundary critical system. We conclude with a brief summary and outlook in Sec.~\ref{sec:outlook}. The appendices contain some technical discussions and proofs. In Appendix \ref{sec:holonomy}, we discuss the possibility of emergent topological order in the bulk using a simple example of a system with $Z_2$ symmetry. In Appendix \ref{app:areaLaw}, we derive the Schmidt decomposition of the bulk state, which can be used to deduce area law entanglement in the bulk and also to construct a gauge-invariant parent Hamiltonian for the bulk state. The latter is described in Appendix \ref{app:parentHam}.

%More recently, it has been argued that the hyperbolic geometry of the MERA can be interpreted as a 1+1 deSitter spacetime \cite{deSitter}, which is a departure from earlier conjectures that it corresponds to the spatial slice of 2+1 anti-deSitter spacetime. We remark that our bulk description does not depend on any specific spacetime interpretation of the MERA's hyperbolic geometry, though a certain ``classical geometry'' may emerge in paticular treatments of our bulk degrees of freedom.

\section{Boundary state with a global symmetry}\label{sec:boundary}
Consider an infinite 1D lattice $\mathcal{L}$ and a compact, completely reducible symmetry group $\myg$. Each site of $\mathcal{L}$ is described by a Hilbert space $\mathbb{V}$ on which the group $\myg$ acts by means of a unitary representation
\begin{equation}
\hat{V}_g : \mathbb{V} \rightarrow \mathbb{V},~~~\hat{V}_g\hat{V}^\dagger_g = \hat{V}^\dagger_g \hat{V}_g = \hat{I},\nonumber
\end{equation}
for all $g \in \mathcal{G}$. Also consider a local Hamiltonian $\hat{H}$ that acts on the lattice $\mathcal{L}$ and has a global symmetry $\myg$, namely,
\begin{equation}
[\hat{H},~\bigotimes_i \hat{V}^{(i)}_g] = 0,~~~\mbox{for all }g \in \myg,
\end{equation}
where $ \hat{V}^{(i)}_g \cong \hat{V}_g$ is a unitary representation of the symmetry group $\myg$ on site $i$.
We assume that the ground state $\mypsi$ of $\hat{H}$ also has a global symmetry $\myg$, namely,
\begin{equation}\label{eq:symstate}
\mypsi = (\bigotimes_i \hat{V}^{(i)}_g)  \mypsi.
\end{equation}
The superscript `bound' appears in anticipation that the ground state plays the role of the boundary state in our holographic correspondence.

%%%%%%%%%%%%%%%%%%%%%%%%%%%%%%%%%%%%%%%%%%%%%%%%%%%%%%%%%%%%%%%%%%%%%%%%%%%%%%%%%%%%%%%%%%%%%%%%%
\begin{figure}[t]
  \includegraphics[width=\columnwidth]{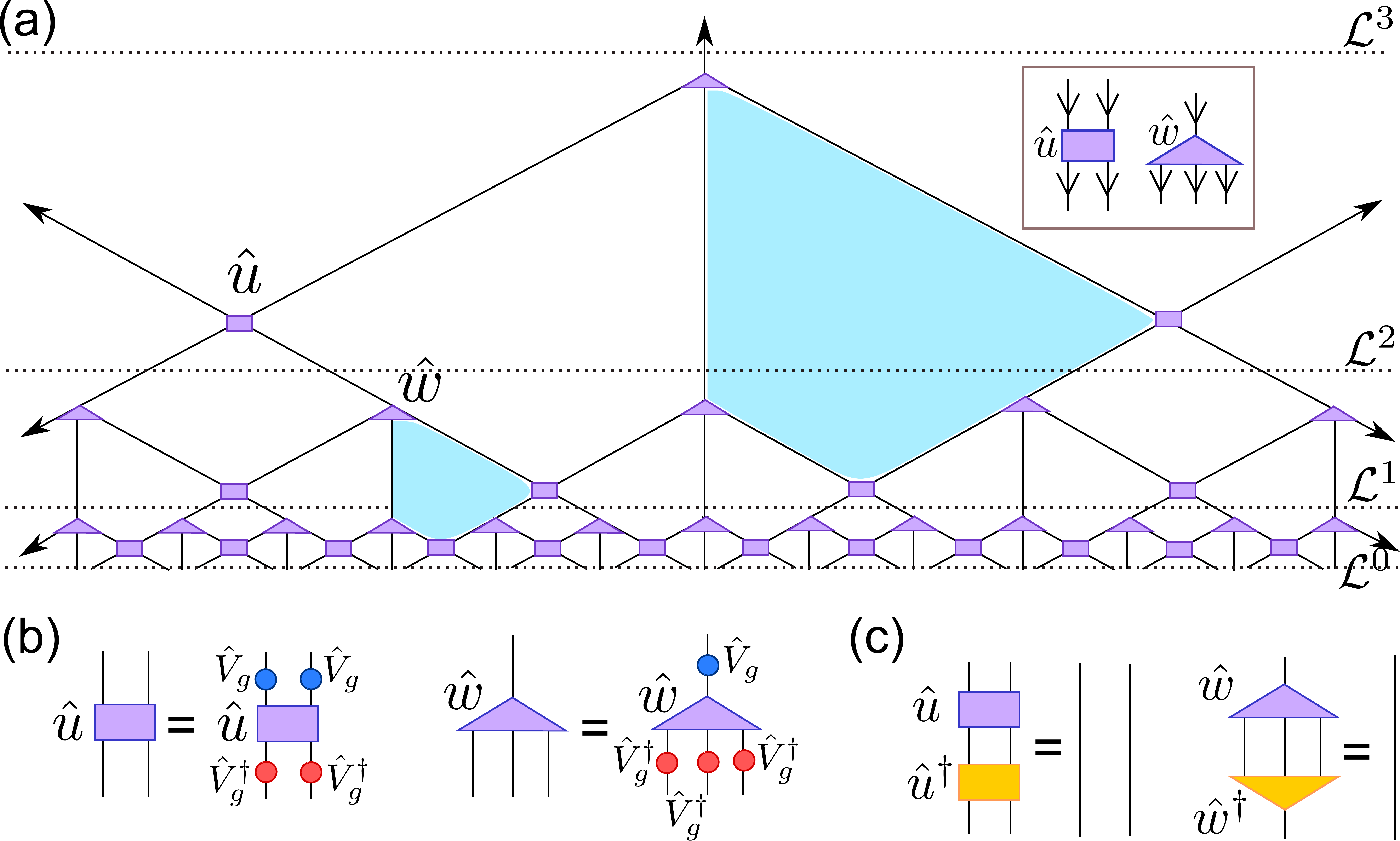}
\caption{\label{fig:mera} (a) Graphical representation of a fragment of the infinite MERA tensor network representation of a quantum many-body state $\mypsi$ of an infinite lattice $\mathcal{L}$. The thick arrows indicate that the tensor network extends infinitely in the top vertical and both horizontal directions. The vertical direction corresponds to length scale; $\mathcal{L}_0 \rightarrow \mathcal{L}_1 \rightarrow \mathcal{L}_2 \rightarrow \mathcal{L}_3\rightarrow \cdots$ is a sequence of increasing coarse-grained lattices where $\mathcal{L}_0 \cong \mathcal{L}$ is the ultraviolet lattice. The MERA may be viewed a tiling of the hyperbolic plane. In the graph metric, in which each edge has unit length, tiles that have the same shape have the same area. For example, the two blue tiles have the same shape but appear to have different areas because we have stretched out a tiling of hyperbolic plane on a flat plane.  (b) Indices are decorated with arrows, as depicted in the box, which indicate how the symmetry acts on the tensors. Tensors $\hat{u}$ and $\hat{w}$ commute with the action of the symmetry as shown, see \eref{eq:meraTensors}. (c) Graphical representation of equalities \eref{eq:isometric} fulfilled by the isometric tensors $\hat{u}$ and $\hat{w}$.}
\end{figure}
%%%%%%%%%%%%%%%%%%%%%%%%%%%%%%%%%%%%%%%%%%%%%%%%%%%%%%%%%%%%%%%%%%%%%%%%%%%%%%%%%%%%%%%%%%%%%%%

In this paper, we represent $\mypsi$ by means of an infinite \textit{symmetry-protected} MERA tensor network.
The tensor network is depicted in \fref{fig:mera}. An \textit{open index} $o_i$ of the MERA labels an orthonormal basis $\{\ket{o_i}\}$ on site $i$ of the lattice $\mathcal{L}$. State $\mypsi$ can be formally expanded as
\begin{equation}
\mypsi = \sum_{o_1,o_2,\ldots} \hat{\Psi}_{o_1,o_2,\ldots} \ket{o_1} \otimes \ket{o_2} \otimes \cdots
\end{equation}
where the probability amplitudes $\hat{\Psi}_{o_1,o_2,\ldots}$ are obtained by contracting the tensor network, which involves summing over all the \textit{bond indices}---indices that connect the tensors in the network.

The MERA representation also describes the RG flow of the ground state. Each layer of tensors of the MERA, separated by dotted lines in \fref{fig:mera}, implements a real space RG transformation---known as \textit{entanglement renormalization}---that maps a lattice $\mathcal{L}^{k}$ with $L~(\rightarrow \infty)$ sites to a coarse-grained lattice $\mathcal{L}^{k+1}$ with $L/3$ sites.
The MERA tensors are chosen so that the renormalization preserves the ground subspace at each step. Subsequent renormalization steps generate a sequence of increasingly coarse-grained lattices: $\mathcal{L}^{0} \rightarrow\mathcal{L}^{1}\rightarrow \mathcal{L}^{2} \cdots$, where $\mathcal{L}^{0} \cong \mathcal{L}$ is the ultraviolet lattice. Thus, the extra dimension of the tensor network corresponds to length scale, in the sense that the residual tensor network obtained by discarding one or more bottom layers is a representation of the ground state on a coarse-grained lattice.

For simplicity, and without loss of generality, in this paper we assume that the ground state $\mypsi$ (and the Hamiltonian $\hat{H}$) is translation invariant and scale-invariant. Specifically, $\mypsi$ is a RG fixed point in a gapped or critical phase.
Subsequently, a MERA representation of $\mypsi$ can be composed from copies of the same two tensors, $\hat{u}$ and $\hat{w}$, throughout the tensor network \cite{MERACFT}, see \fref{fig:mera}.

We decorate the indices of the MERA tensors with arrows, as depicted in \fref{fig:mera}(a), which indicate how the symmetry acts on the tensors.
Tensors $\hat{u}$ and $\hat{w}$ are linear transformations from input spaces (incoming indices) to output spaces (outgoing indices) as $\hat{u}:\mathbb{V} \otimes \mathbb{V} \rightarrow \mathbb{V} \otimes \mathbb{V}$ and $\hat{w} : \mathbb{V} \rightarrow \mathbb{V} \otimes \mathbb{V} \otimes \mathbb{V}$. The symmetry acts as $\hat{V}_g$ on an incoming index (input space) and as $\hat{V}^\dagger_g$ on an outgoing index (output space). Tensors $\hat{u}$ and $\hat{w}$ remain invariant under the action of the symmetry, namely,
\begin{equation}\label{eq:meraTensors}
\begin{split}
\hat{u} &= (\hat{V}_g \otimes \hat{V}_g)~ \hat{u}~ (\hat{V}^\dagger_g \otimes \hat{V}^\dagger_g), \\
\hat{w} &= (\hat{V}_g)~\hat{w}~(\hat{V}^\dagger_g \otimes \hat{V}^\dagger_g \otimes \hat{V}^\dagger_g),
\end{split}
\end{equation}
for all group elements $g \in \mathcal{G}$, as depicted in \fref{fig:mera}(b). For brevity, we say that tensors $\hat{u}$ and $\hat{w}$ are $\mathcal{G}$\textit{-symmetric}. The choice of $\mathcal{G}$-symmetric tensors captures the global symmetry, \eref{eq:symstate}, exactly and also generates a symmetry protected RG flow \cite{Singh131,Singh}.
The tensors $\hat{u}$ and $\hat{w}$ are also isometric, namely, they satisfy
\begin{equation}\label{eq:isometric}
\sum_{rs} (\hat{u})^{pq}_{rs}(\hat{u}^\dagger)^{rs}_{p'q'} = \delta^p_{p'}\delta^q_{q'},~\sum_{qrs} (\hat{w})^{p}_{qrs}(\hat{w}^\dagger)^{qrs}_{p'} = \delta^p_{p'},
\end{equation}
depicted in \fref{fig:mera}(c). 

%%%%%%%%%%%%%%%%%%%%%%%%%%%%%%%%%%%%%%%%%%%%%%%%%%%%%%%%%%%%%%%%%%%%%%%%%%%%%%%%%%%%%%%%%%%%%%%%%%
\begin{figure}
  \includegraphics[width=\columnwidth]{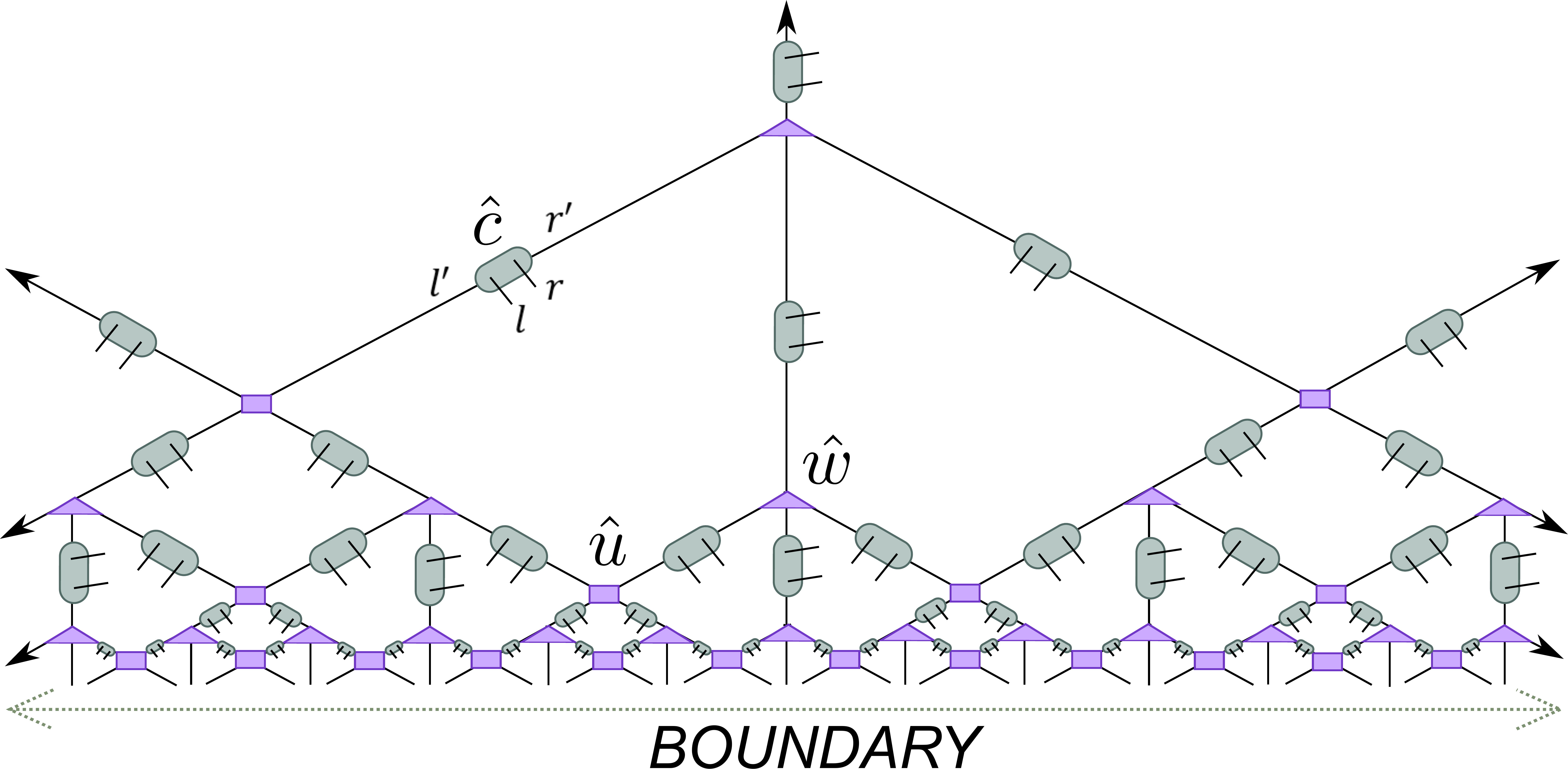}
\caption{\label{fig:lift} The \textit{lifted MERA} tensor network obtained by inserting a 4-index tensor $(\hat{c})^{lr}_{l'r'}$ on every bond of the MERA representation of a 1D quantum many-body state. Each open index of the lifted MERA labels an orthonormal basis on a different site of the bulk lattice $\mathcal{M}$. Each bond of the MERA is associated with two bulk sites, corresponding to the open indices $l$ and $r$. These two sites carry the `left' and `right' gauge transformations respectively.  The lifted MERA represents a quantum state of $\mathcal{M}$, whose probability amplitudes are obtained by contracting all the tensors of the lifted tensor network. The tensor $\hat{c}$ is not fixed and parameterizes our ansatz for the holographic dual of the 1D state.
%The lifted tensor network continues infinitely in all directions, though in practice one can impose a boundary on the geometry corrresponding to choosing a ultra-violet cutoff in the boundary description, see Ref.\onlinecite{TNC}. In this paper, we will impose a UV cut-off in the boundary description, which corresponds to the appearance of open indices at the boundary of the lifted MERA.
}
\end{figure}
%%%%%%%%%%%%%%%%%%%%%%%%%%%%%%%%%%%%%%%%%%%%%%%%%%%%%%%%%%%%%%%%%%%%%%%%%%%%%%%%%%%%%%%%%%%%%%%%%%

\section{Dual bulk state}\label{sec:bulk}
In this section, we introduce a holographic description of the 1D state $\mypsi$ by extending the construction presented in Ref.~\onlinecite{TNC} to the presence of symmetries. We refer the reader to Ref.~\onlinecite{TNC} for a discussion about how the construction is inspired by and implements certain general features of the AdS/CFT correspondence.

Let us embed the MERA in a 2D manifold with a boundary, such that the open indices of the MERA are located at the boundary of the manifold and all the bond indices are located inside the bulk of the manifold.
%We then view the MERA as tiling the plane. Consequently, the graph metric underlying the MERA induces a hyperbolic geometry on the plane, see \fref{fig:mera}.  For example, in the graph metric (where each edge has unit length) the two shaded tiles have the same size but here appear to have different sizes because we have stretched out a tiling of hyeprbolic plane on a flat plane.
Construct a 2D lattice $\mathcal{M}$ on the manifold by locating two sites---each of which is described by the vector space $\mathbb{V}$---on every bond of the tensor network. Lattice $\mathcal{M}$ is simply a collation of the degrees of freedom that appear in the RG flow of the ground state $\mypsi$, and inherits the hyperbolic geometry of the tensor network.

%%%%%%%%%%%%%%%%%%%%%%%%%%%%%%%%%%%%%%%%%%%%%%%%%%%%%%%%%%%%%%%%%%%%%%%%%%%%%%%%%%%%%%%%%%%%%%%%%%
\begin{figure}[b]
  \includegraphics[width=8cm]{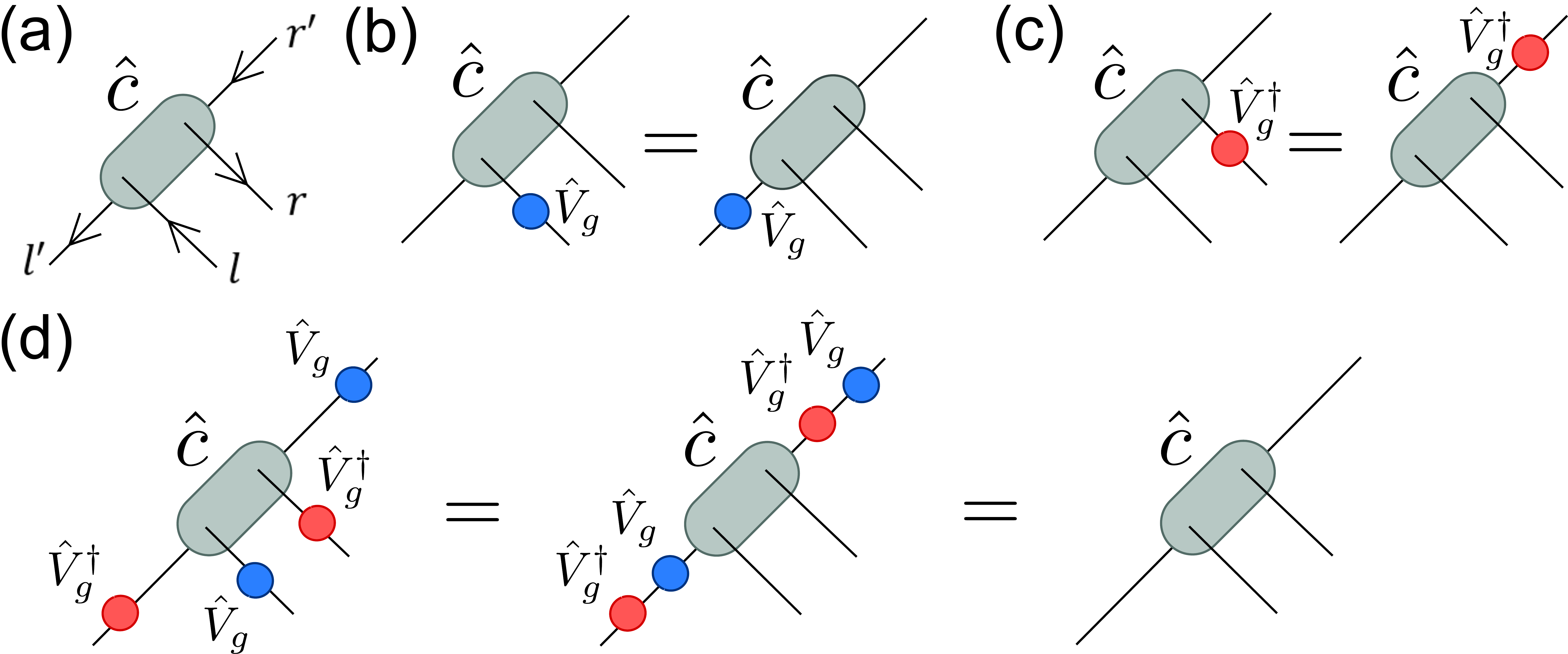}
\caption{\label{fig:copy} (a) Graphical representation of the copy tensor $(\hat{c})^{lr}_{l'r'}$. The symmetry acts as $\hat{V}_g$ (blue solid circle) on an incoming index and as $\hat{V}^\dagger_g$ (red solid circle) on an outgoing index for all $g \in \mathcal{G}$. (b,c) The action of a symmetry operator on index $l$ (index $r$) transfers to the index $l'$ (index $r'$). The left hand side of each equality depicts the action of the symmetry on the copy tensor according to index arrows, while the right hand side depicts an equivalent action of the symmetry on the tensor (not necessarily according to the arrows). (d) Tensor $\hat{c}$ remains invariant under the action of the symmetry according to the index arrows.}
\end{figure}
%%%%%%%%%%%%%%%%%%%%%%%%%%%%%%%%%%%%%%%%%%%%%%%%%%%%%%%%%%%%%%%%%%%%%%%%%%%%%%%%%%%%%%%%%%%%%%%%%%

%%%%%%%%%%%%%%%%%%%%%%%%%%%%%%%%%%%%%%%%%%%%%%%%%%%%%%%%%%%%%%%%%%%%%%%%%%%%%%%%%%%%%%%%%%%%%%%%%
\begin{figure}[t]
  \includegraphics[width=\columnwidth]{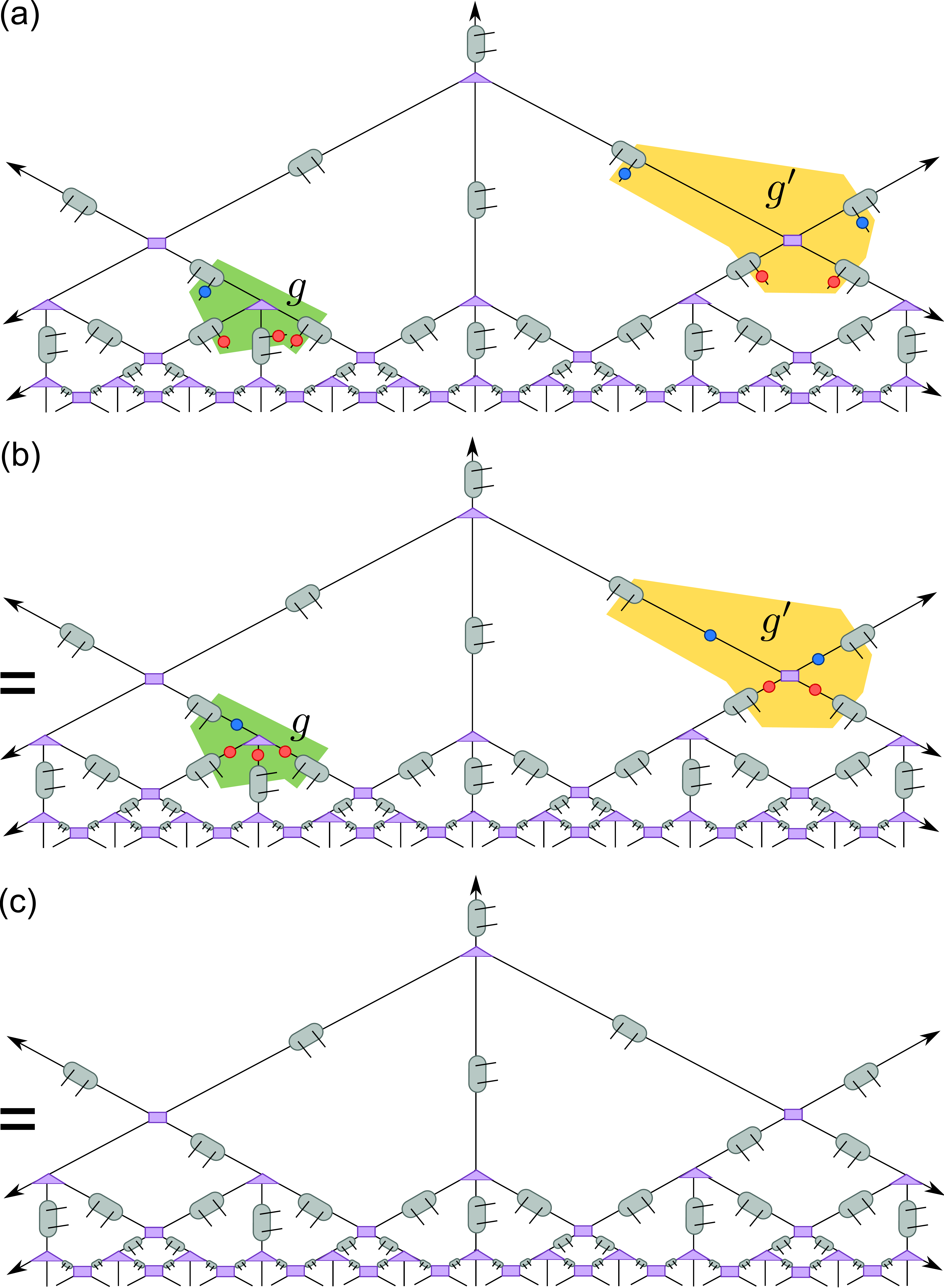}
\caption{\label{fig:gauge} Local gauge symmetry of the bulk state $\myphi$. Here we illustrate that the lifted MERA, and thus $\myphi$, remains invariant under the action of two elementary gauge transformations on the bulk lattice $\mathcal{M}$, corresponding to two different group elements $g,g' \in \mathcal{G}$ respectively. One gauge transformation acts on the 4 bulk sites (counting clockwise starting at the top left in the graphical representation) immediately surrounding tensor $\hat{w}$ (highlighted green) as $\hat{V}_g \otimes \hat{V}^\dagger_g \otimes \hat{V}^\dagger_g \otimes \hat{V}^\dagger_g$, and the other acts on the 4 bulk sites immediately surrounding tensor $\hat{u}$  (higlighted yellow) as
$\hat{V}_{g'} \otimes \hat{V}_{g'} \otimes \hat{V}^\dagger_{g'} \otimes \hat{V}^\dagger_{g'}$. This is shown by means of two equalities: (a)=(b), which results from the symmetry properties of the copy tensor [\fref{fig:copy}$(b)$-$(d)$], and (b)=(c), which results from the fact that the tensors $\hat{u}$ and $\hat{w}$ are $\mathcal{G}$-symmetric [\fref{fig:mera}$(b)$].}
\end{figure}
%%%%%%%%%%%%%%%%%%%%%%%%%%%%%%%%%%%%%%%%%%%%%%%%%%%%%%%%%%%%%%%%%%%%%%%%%%%%%%%%%%%%%%%%%%%%%%%%%

Next, let us insert the 4-index \textit{copy tensor} $(\hat{c})^{lr}_{l'r'}$ on each bond of the MERA, as depicted in \fref{fig:lift}, such that indices $l$ and $r$ are left open. We will define the components of the copy tensor in the next section (see Eqs.~\ref{eq:bondtensor}-\ref{eq:copycomponents}), but here it suffices to say that, colloquially, tensor $\hat{c}$ copies the basis states on a bond index of the MERA to each of the two open indices $l$ and $r$. These indices label an orthonormal basis on the two sites located on that bond respectively, and in analogy to lattice gauge theory, we require that these indices carry the left and right gauge transformations on the bulk lattice $\mathcal{M}$ respectively (as described in Sec.~\ref{ssec:gaugesym}). To this end, we demand that the copy tensor $\hat{c}$ fulfill the following equations that involve the action of symmetry on a single index of the tensor:
\begin{equation}\label{eq:bondsym}
\begin{split}
\sum_{x} (\hat{V}_{g})_{x}^{l}~(\hat{c})^{xr}_{l'r'} &= \sum_{x} (\hat{V}_{g})_{l'}^{x}~ (\hat{c})^{lr}_{xr'},\\
\sum_{x} (\hat{V}_{g}^\dagger)_{x}^{r}~(\hat{c})^{lx}_{l'r'} &= \sum_{x} (\hat{V}_{g}^\dagger)_{r'}^{x}~ (\hat{c})^{lr}_{l'x},
\end{split}
\end{equation}
see \fref{fig:copy}.
%These equations do not completely fix the bond tensor $\hat{c}$. Thus, the lifted MERA describes a class of states on the lattice $\mathcal{M}$, which is our ansatz for the bulk state dual to $\mypsi$.

The new tensor network---the MERA with a copy tensor inserted on every bond---can be viewed as a representation of a quantum state $\myphi$ of the bulk lattice $\mathcal{M}$, where the probability amplitudes of $\myphi$ are (formally) obtained by contracting all its bond indices, analogous to how the MERA encodes the state $\mypsi$. Thus, we have `lifted' the MERA representation of a quantum state $\mypsi$ of the 1D lattice $\mathcal{L}$ to a quantum state $\myphi$ of the 2D lattice $\mathcal{M}$.  We refer to the bulk tensor network, comprised of copies of the ground state tensors $\hat{u}, \hat{w}$ and the copy tensor $\hat{c}$, as the \textit{lifted MERA}.
%Our goal is to determine states within this ansatz that capture some general features of the AdS/CFT correspondence \cite{TNC}. In this paper, we focus on the holographic translation between the boundary and bulk symmetries: 

%Notice that the bulk description (the lifted MERA) also consists of degrees of freedom that appear at the boundary of $\mathcal{M}$, which are associated with the open indices $o_1,o_2, \cdots$ of the lifted MERA. These degrees of freedom are in one to one correspondence with the sites of the lattice $\mathcal{L}$. Depending on how the boundary of the two dimensional geometry is treated, these boundary degrees of freedom can be regarded either as degrees of freedom that appear at an asymptotic boundary or those resulting from the choice of an ultra-violet cut off length scale in the boundary description \cite{TNC}.
% If $\hat{c}$ is the copy tensor, $\hat{c}^{\mbox{\tiny deg}}$ can be decomposed in terms of two $\hat{x}$ tensors, \eref{eq:copycomponents}. (b) A useful equality satisfed by the copy tensor. 

%%%%%%%%%%%%
\subsection{Local gauge symmetry}\label{ssec:gaugesym}
An immediate consequence of the symmetry conditions \eref{eq:bondsym} is that the bulk state $\myphi$ has a local gauge symmetry $\mathcal{G}$. Let us introduce gauge transformations on the bulk lattice $\mathcal{M}$ as follows.
The symmetry $\mathcal{G}$ acts on the two sites located on a bond differently, namely, as $\hat{V}_g$ and $\hat{V}^\dagger_g$ respectively. (This choice corresponds to the action of `left' and `right' gauge transformations in lattice gauge theory.)
Elementary gauge transformations act on the 4 bulk sites (counting clockwise starting at the top left in the graphical representation) immediately surrounding tensor $\hat{u}$ as
$\hat{V}_g \otimes \hat{V}_g \otimes \hat{V}^\dagger_g \otimes \hat{V}^\dagger_g$,
and on the 4 bulk sites immediately surrounding tensor $\hat{w}$ as 
$\hat{V}_g \otimes \hat{V}^\dagger_g \otimes \hat{V}^\dagger_g \otimes \hat{V}^\dagger_g$.
General gauge transformations act on larger regions of the lattice $\mathcal{M}$ by composing these elementary gauge transformations.

Let us consider the result of applying an elementary gauge transformation on a dual bulk state $\myphi$ that is represented by a lifted MERA. In the lifted tensor network representation, the action of an elementary gauge transformation corresponds to contracting the symmetry operators on the open indices $l,r$ of the bond tensors located immediately around an $\hat{u}$ or $\hat{w}$ tensor. The symmetry acts as $\hat{V}_g$ and $\hat{V}^\dagger_g$ on the open indices $l$ (`left') and $r$ (`right') respectively. Owing to \eref{eq:bondsym} [\fref{fig:copy}], operator $\hat{V}_g$ that is applied on an open index of a copy tensor `slides' through to a bond index of the lifted MERA. Consequently, the action of a gauge transformation on the bulk state translates to contracting the symmetry operators with the tensors around which they are applied, see \fref{fig:gauge}. However, the tensors are $\mathcal{G}$-symmetric [\eref{eq:meraTensors} and \fref{fig:mera}(b)], which eliminates the symmetry operators. Thus, the lifted MERA, and therefore state $\myphi$, remains invariant under the action of local gauge transformations.

The gauge invariance of the bulk state is manifest in the same way as appears in lattice gauge theory as originally formulated by Kogut. There one introduces basis states on edges with the local degrees of freedom split into three subspaces as $\ket{j,n_L,n_R}$ where $j$ labels an irrep of the group, and the other two are labelled by matrix elements of these representations (one on the left and one on the right side of the edge). This is a Fourier basis conjugate to the group element labelled basis the two bases being related by the Peter-Weyl theorem. As in lattice gauge theory, the physical states in the bulk are those that are invariant under gauge transformations on a vertex that act with the same group element on all the neighbouring carrier spaces on the edges incident to that site. For Abelian models, all irreps are one dimensional so only one bulk degree of freedom would be needed per edge, but for non-Abelian gauge groups, two labels are needed since the irreps are matrices with components labelled by left and right pairs. We also remark that Elitzur’s theorem applies to our bulk state in the sense that the expectation value of non-gauge invariant quantities are trivial by construction.

\section{Spin network decomposition of bulk states}\label{sec:spinnetwork}

Let us now introduce a basis in the vector space $\mathbb{V}$, which describes each site of the boundary lattice $\mathcal{L}$ and also each site of the bulk lattice $\mathcal{M}$. Under the action of the symmetry, $\mathbb{V}$ generally decomposes as
\begin{equation}\label{eq:onesite}
\mathbb{V} \cong \bigoplus_j \mathbb{D}_{j} \otimes \mathbb{S}_j,
\end{equation}
where the symmetry acts on space $\mathbb{S}_j$ by means of the irreducible representation (irrep) of $\mathcal{G}$ labelled by quantum number (or charge) $j$, and $\mathbb{D}_{j}$ is the \textit{degeneracy} space of irrep $j$. Accordingly, the symmetry operators $\hat{V}_g$ decompose as
\begin{equation}\label{eq:symDecompose}
\hat{V}_g = \bigoplus_j (\hat{I}_{d_j} \otimes  \hat{V}_{g,j}),~~~\forall g \in \mathcal{G}.
\end{equation}
In particular, note that the symmetry operators act trivially on the degeneracy spaces, namely, as the $d_j \times d_j$ identity $\hat{I}_{d_j}$ on the degeneracy space $\mathbb{D}_{j}$, where $d_j$ is the dimension of the space $\mathbb{D}_{j}$.

We denote by $\{\ket{j,m_j}\}$ an orthonormal basis in the irrep space $\mathbb{S}_j$, by $\{\ket{j,t_j}\}$ an orthonormal basis in the degeneracy space $\mathbb{D}_{j}$, and by $\{\ket{j,t_j,m_j} \equiv \ket{j,t_j} \otimes \ket{j,m_j}\}$ the basis on the total space $\mathbb{V}$. For example, if $\myg = SU(2)$, then the symmetry charge $j \in \{0,\frac{1}{2},1,\frac{3}{2},\cdots\}$ is the total spin, $m\in\{-j,-j+1,\cdots,j\}$ is the spin projection along the $z$-axis. (For simplicity, we assume that $\myg$ is multiplicity-free.)
%The presence of multiplicities requires an additional label for the basis.
The description simplifies considerably for an Abelian symmetry, for example $\myg = Z_n, U(1)$, since all the irreps of an Abelian group have dimension 1, that is, dim($S_j$)=1 for all $j$.

%Wigner-Eckart decompositions of the various $\mathcal{G}$-symmetric tensors in the irrep basis \eref{eq:symBasis}. 
 %%%%%%%%%%%%%%%%%%%%%%%%%%%%%%%%%%%%%%%%%%%%%%%%%%%%%%%%%%%%%%%%%%%%%%%%%%%%%%%%%%%%%%%%%%%%%%%%%
\begin{figure}[t]
  \includegraphics[width=\columnwidth]{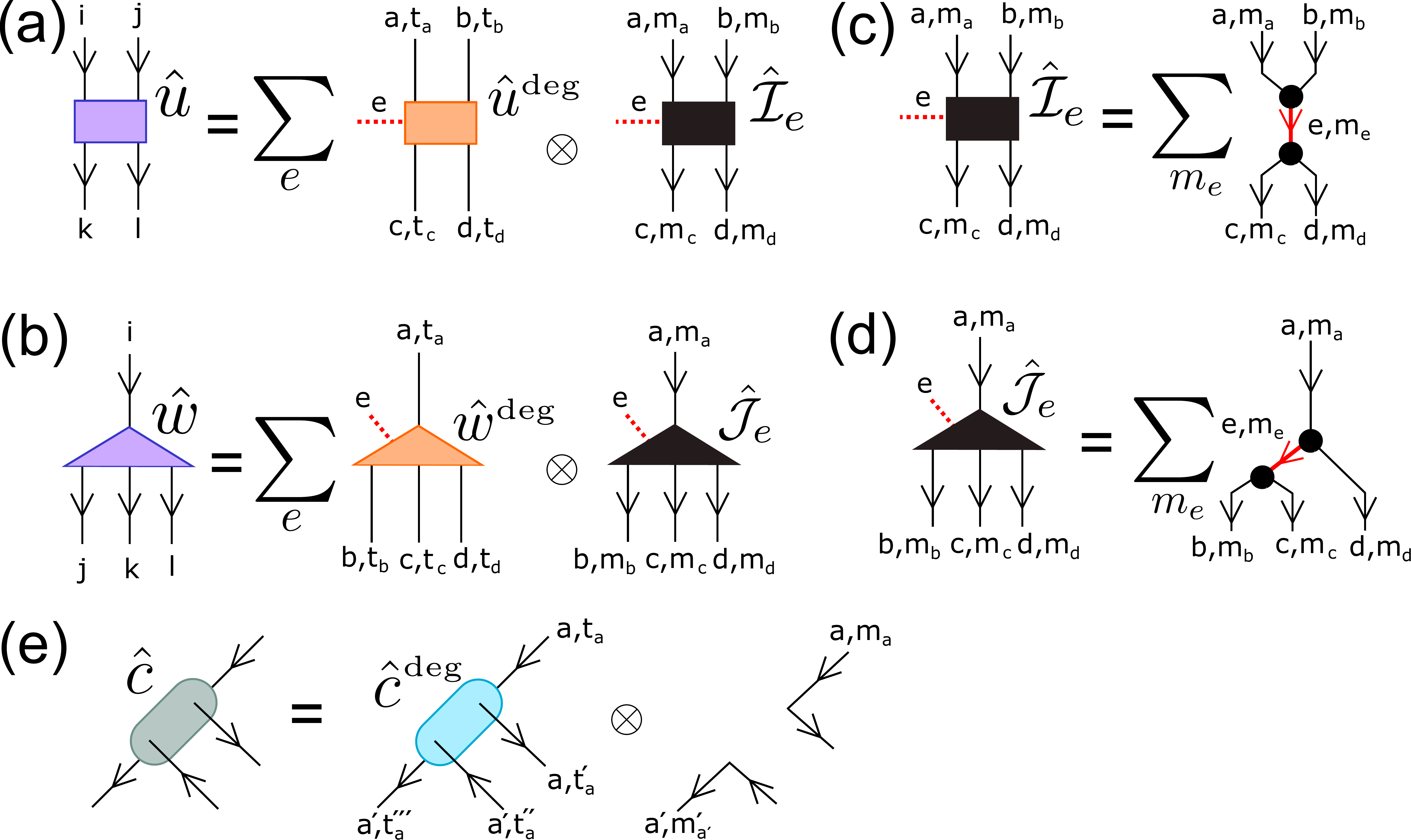}
\caption{\label{fig:symTensors} (a,b) Wigner-Eckart decomposition of tensors $\hat{u}$ and $\hat{w}$ into degeneracy tensors and intertwiners of the symmetry group $\mathcal{G}$, \eref{eq:symtTenDecom}. (c,d) Intertwiners $\hat{\mathcal{I}}_e$ and $\hat{\mathcal{J}}_e$ expressed in terms of two Clebsch-Gordan coefficients (solid black circles), \eref{eq:int1} and \eref{eq:int2}. The red lines carry the intermediate intertwining charges $e$. (e) Wigner-Eckart decomposition of the copy tensor $\hat{c}$ according to \eref{eq:bondtensor}. The symmetry properties depicted in \fref{fig:copy}$(b)$-$(c)$ imply that the intermediate intermediate charge $e$ is trivial, $e=0$ (depicted by the absence of any red line).} 
\end{figure}
%%%%%%%%%%%%%%%%%%%%%%%%%%%%%%%%%%%%%%%%%%%%%%%%%%%%%%%%%%%%%%%%%%%%%%%%%%%%%%%%%%%%%%%%%%%%%%%

 %%%%%%%%%%%%%%%%%%%%%%%%%%%%%%%%%%%%%%%%%%%%%%%%%%%%%%%%%%%%%%%%%%%%%%%%%%%%%%%%%%%%%%%%%%%%%%%%%
\begin{figure*}[t]
  \includegraphics[width=18.5cm]{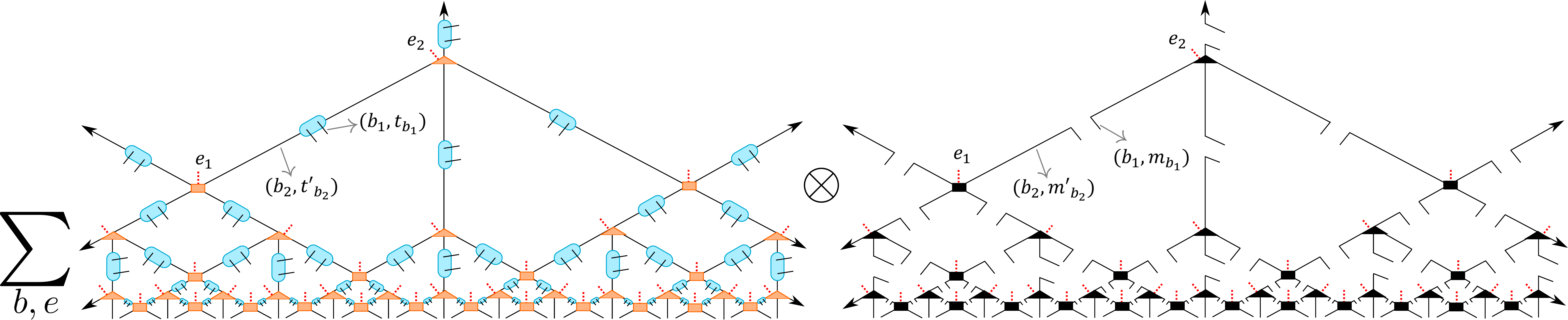}
\caption{\label{fig:spinnetwork} The lifted MERA decomposes as a sum of tensor product of two parts: (left)  a tensor network composed of degeneracy tensors, and  (right) a tensor network composed of intertwiners of the symmetry group, namely, a spin network. The sum is over tuples of symmetry charges $b$ and $e$. Here $b \equiv (b_1,b_2,\cdots)$ is the tuple of symmetry charges carried by all bond and open indices of the lifted MERA, and $e\equiv (e_1,e_2,\cdots)$ is the tuple of \textit{intertwining} symmetry charges associated with all the intertwiners in the spin network, see \fref{fig:symTensors}. The decomposition separates out the gauge degrees of freedom, dual to the global symmetry at boundary, from the remaining bulk degrees of freedom. The spin network states span the gauge-invariant support of the bulk state, while we view the remaining degrees of freedom to possibly include `gravitational' degrees of freedom in a holographic interpretation of the MERA.}
\end{figure*}
%%%%%%%%%%%%%%%%%%%%%%%%%%%%%%%%%%%%%%%%%%%%%%%%%%%%%%%%%%%%%%%%%%%%%%%%%%%%%%%%%%%%%%%%%%%%%%%

According to the Wigner-Eckart theorem, the $\myg$-symmetric tensors $\hat{u}$ and $\hat{w}$ [\eref{eq:meraTensors}] decompose in terms of the intertwiners of $\myg$. If the components of tensors $\hat{u}$ and $\hat{w}$ are denoted as $(\hat{u})^{pq}_{rs}$ and $(\hat{w})^{p}_{qrs}$ respectively, then in the irrep basis
\begin{equation}\label{eq:symBasis}
\begin{split}
\ket{p} &\equiv \ket{a,t_a,m_a},~~~\ket{q} \equiv \ket{b,t_b,m_b},\\ 
\ket{r} &\equiv \ket{c,t_c,m_c},~~~~ \ket{s} \equiv \ket{d,t_d,m_d},
\end{split}
\end{equation}
where $a,b,c,d$ denote symmetry charges, the tensors decompose as
\begin{equation} \label{eq:symtTenDecom}
\begin{split}
\hat{u} &\equiv \bigoplus_{abcd;e}~(\hat{u}^{\mbox{\tiny deg}}_e)^{ab}_{cd} \otimes (\hat{\mathcal{I}}_e)^{ab}_{cd},\\
\hat{w} &\equiv \bigoplus_{abcd;f}~(\hat{w}^{\mbox{\tiny deg}}_f)^{a}_{bcd} \otimes (\hat{\mathcal{J}}_f)^{a}_{bcd},
\end{split}
\end{equation}
depicted in \fref{fig:symTensors}(a)-(b).
Here
\begin{equation}
\begin{split}
(\hat{\mathcal{I}}_e)^{ab}_{cd} &: (\mathbb{S}_a \otimes \mathbb{S}_b) \rightarrow (\mathbb{S}_c \otimes \mathbb{S}_d) \\
(\hat{\mathcal{J}}_f)^{a}_{bcd} &: \mathbb{S}_a \rightarrow (\mathbb{S}_b \otimes \mathbb{S}_c \otimes \mathbb{S}_d) \nonumber
\end{split}
\end{equation}
are 4-index intertwiners of the symmetry group $\mathcal{G}$, whose components are completely fixed by the properties of the group representations. (The intermediate charges $e$ and $f$ label a basis in a vector space of intertwiners.) The components of $(\hat{\mathcal{I}}_e)^{ab}_{cd}$ are given by [see \fref{fig:symTensors}(c)]
\begin{equation}\label{eq:int1}
\begin{split}
[(\hat{\mathcal{I}}_e)^{ab}_{cd}]^{m_am_b}_{m_cm_d} \equiv  \sum_{m_e} &\braket{e,m_e}{a,m_a;b,m_b}\\
&\braket{e,m_e}{c,m_c;d,m_d},
\end{split}
\end{equation}
where, for example, $\braket{e,m_e}{a,m_a;b,m_b} \equiv \bra{e,m_e}\cdot(\ket{a,m_a}\otimes \ket{b,m_b})$ are the Clebsch-Gordan coefficients that describe the change of basis from the tensor product basis $\ket{a,m_a}\otimes \ket{b,m_b}$ to the total charge basis $\ket{e,m_e}$. Analogously, we have [see \fref{fig:symTensors}(d)]
\begin{equation}\label{eq:int2}
\begin{split}
[(\hat{\mathcal{J}}_f)^{a}_{bcd}]^{m_a}_{m_bm_cm_d} \equiv \sum_{m_f} &\braket{f,m_f}{b,m_b;cm_c}\\
&\braket{a,m_a}{f,m_f;dm_d}.
\end{split}
\end{equation}
Finally, $(\hat{u}^{\mbox{\tiny deg}}_e)^{ab}_{cd}$ and $(\hat{w}^{\mbox{\tiny deg}}_f)^{a}_{bcd}$ in \eref{eq:symtTenDecom} are \textit{degeneracy tensors}, namely, multi-linear maps between the degeneracy spaces,
\begin{equation}
\begin{split}
(\hat{u}^{\mbox{\tiny deg}}_e)^{ab}_{cd} &: (\mathbb{D}_a \otimes \mathbb{D}_b) \rightarrow (\mathbb{D}_c \otimes \mathbb{D}_d) \\
(\hat{w}^{\mbox{\tiny deg}}_f)^{a}_{bcd} &: \mathbb{D}_a \rightarrow (\mathbb{D}_b \otimes \mathbb{D}_c \otimes \mathbb{D}_d), \nonumber
\end{split}
\end{equation}
and represent the part of the tensor that is not fixed by the symmetry. We refer the reader to Refs~\cite{Singh} for a more detailed exposition on such decompositions of $\myg$-symmetric tensors.

The copy tensor $\hat{c}$ is $\mathcal{G}$-symmetric [\fref{fig:copy}(d)] and therefore also decomposes according to the Wigner-Eckart theorem. The equalities \eref{eq:bondsym} imply that only the \textit{trivial} interwiner appears in the decomposition, namely, an intertwiner with trivial intermediate charge. E.g., for $\mathcal{G} = SU(2)$ the trivial charge corresponds to the spin 0 irrep.
Specifically, tensor $\hat{c}$ decomposes as (see \fref{fig:symTensors}(e))
\begin{equation}\label{eq:bondtensor}
\hat{c} \equiv \bigoplus_{a,a'}~\hat{c}^{\mbox{\tiny deg}}_{a,a'} \otimes (\hat{\mathcal{I}}_0)^{aa^*}_{a'{a'}^*},
\end{equation}
where $\hat{c}^{\mbox{\tiny deg}}_{a,a'}$ is a 4-index degeneracy tensor, $a^*$ denotes the conjugate charge of $a$ (namely, charges $a$ and $a^*$ fuse to the trivial charge), and $\hat{\mathcal{I}}_0$ is the intertwiner defined according to \eref{eq:int1} for the trivial intermediate charge $e=0$. The intertwiner $\hat{\mathcal{I}}_0$ is, in fact, equal to the tensor product of the identity $\hat{I}_a$ and the identity $\hat{I}_{a'}$ on the irrep spaces $\mathbb{S}_a$ and $\mathbb{S}_{a'}$ respectively [\eref{eq:onesite}]. 

Denote the components of a degeneracy tensor $\hat{c}^{\mbox{\tiny deg}}_{a,a'}$ by $(\hat{c}^{\mbox{\tiny deg}}_{a,a'})^{t_at'_{a}}_{\tilde{t}_{a'}\tilde{t}'_{a'}}$ where $t_a,t'_a,\tilde{t}_{a'},\tilde{t}'_{a'} \in \{1,2,\ldots,d_a\}$. The only non-zero components are given by
\begin{equation}\label{eq:copycomponents}
(\hat{c}^{\mbox{\tiny deg}}_{a,a})^{t_at_a}_{t_at_a}=1.
\end{equation}
Equations~(\ref{eq:bondtensor})-(\ref{eq:copycomponents}) define the copy tensor.

%The degeneracy tensors $\{\hat{c}^{\mbox{\tiny deg}}_{a,a'}\}$, which are not constrained by the symmetry, define our bulk ansatz.
By decomposing tensors according to \eref{eq:symtTenDecom} and \eref{eq:bondtensor}, the entire lifted MERA tensor network decomposes as shown in \fref{fig:spinnetwork}. Here the sum is over the symmetry charges carried by all the indices of the tensor network, and also the internal intertwining charges that appear in the decomposition of each tensor. The tensor networks appearing on the left in the figure are composed only of the degeneracy tensors, and represent the support of the bulk state $\myphi$ on the sector of the Hilbert space that is not constrained by the symmetry. On the other hand, the tensor networks appearing on the right in \fref{fig:spinnetwork} are composed only of intertwiners of $\mathcal{G}$, and are thus completely fixed by the symmetry. These tensor networks are nothing but \textit{spin network states}, which here label an orthonormal basis in the support of the bulk state within the gauge-invariant subspace of the bulk lattice $\mathcal{M}$, analogous to their role in lattice gauge theories \cite{BaezSpinNetwork}.

In order to make an analogy with the AdS/CFT correspondence, we interpret the degeneracy degrees of freedom as possibly including `gravitational' degrees of freedom. (Or more generally, `emergent' gauge degrees of freedom, see Sec. \ref{ssec:emergent}.) Thus, in the context of holography, the bulk state $\myphi$ may be interpreted as an entangled state of gauge fields (described by the spin networks) living on a 2D quantum geometry (described by the degeneracy tensors). We remark that the bulk construction described here may be readily generalized by also exposing and lifting the internal intertwining charges (that is, the $e_i$'s that appear in \fref{fig:spinnetwork} also appear as open indices in the lifted MERA), which allows to incorporate `gauge matter' in the model. However, we do not pursue this here.

We remark that in a lattice gauge theory, based on a continous gauge group $\mathcal{G}$, one often has to truncate the irreps that appear on the bonds of the spin networks, in order to make calculations tractable. In our bulk construction, the irreps that appear on the bonds of the holographic spin networks are also truncated, since they are carried over from the MERA representation of the ground state. However, the trucation here results from practical considerations in MERA simulations. One systematically assigns only a finite number of irreps on the bonds of the MERA in the variational energy minimization for a given $\mathcal{G}$-symmetric Hamiltonian. Bond irreps are selected with the aim of obtaining the smallest energy possible, within the constraints imposed by the available computational resources.

The spin network decomposition separates the gauge degrees of freedom from the remaining (degeneracy) degrees of freedom in the bulk. This leads to three interesting applications.
First, the decomposition allows one to introduce meaningful gauge-invariant observables in the bulk, since it exposes quantum numbers in the bulk (within the gauge-invariant sector). Second, it reveals correlations between the gauge and the remaining degrees of freedom. And third, it allows one to trace out the gauge degrees of freedom and thus probe the nature of the degeneracy degrees of freedom. We make some remarks pertaining to the first application in Sec.~\ref{ssec:gaugebulkops} below. The second and third applications are explored in Sec.~\ref{sec:bulkentanglement}.
\begin{figure*}[t]
  \includegraphics[width=18cm]{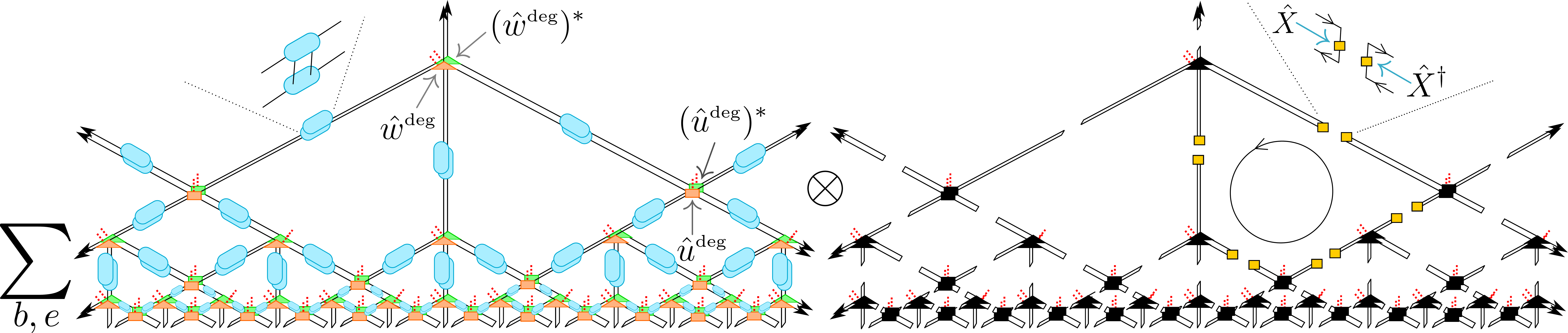}
\caption{\label{fig:holonomy} An illustration of the tensor network contraction equating to the expectation value of a gauge-invariant loop operator in the bulk that acts non-trivially on the gauge degrees of freedom (the spin networks) and as the identity on the remaining degrees of freedom. For $\myg = Z_2$ and $\hat{X}$ defined according to \eref{eq:XX}, the loop operator can be understood as a Wilson loop in a $Z_2$ lattice gauge theory (here defined on a hyperbolic lattice).}
\end{figure*}
%%%%%%%%%%%%%%%%%%%%%%%%%%%%%%%%%%%%%%%%%%%%%%%%%%%%%%%%%%%%%%%%%%%%%%%%%%%%%%%%%%%%%%%%%%%%%%%

\subsection{Gauge-invariant bulk operators}\label{ssec:gaugebulkops}
As mentioned above, the spin network decomposition of the lifted MERA allows one to introduce gauge-invariant operators in the bulk. Simple examples are operators that act non-trivially on the spin network states and as the identity on the degeneracy degrees of freedom.  Figure \ref{fig:holonomy} illustrates a tensor network contraction that equates to the expectation value of such a gauge-invariant (wilson) loop operator in the bulk. See Appendix \ref{app:holonomy} for examples of interesting gauge-invariant loop operators. In the context of holography, it may also be possible to infer some information about the curvature of the ambient space in which the gauge field lives \cite{CurvedHolonomy}. For a pure gauge theory on a flat space the vacuum state is described as having a flat connection everywhere.
However, for curved space, the expectation value will differ in general. Thus, we could expect to infer metric curvature by measuring local holonomies. 

%In the context of holography, it may also be possible to infer some information about the curvature of the ambient space in which the gauge field lives from the wilson loops
Gauge-invariant loop operators may also be used to detect topological order in the bulk. In Appendix \ref{sec:holonomy}, we explore the topological order of the bulk state for the simple case of $Z_2$ symmetry. (The discussion readily generalizes to $Z_n$ symmetry.)

%%%%%%%%%%%%%%%%%%%%%%%%%%%%%%%%%%%%%%%%%%%%%%%%%%%%%%%%%%%%%%%%%%%%%%%%%%%%%%%%%%%%%%%%%%%%%%%%%
\begin{figure}[b]
  \includegraphics[width=7cm]{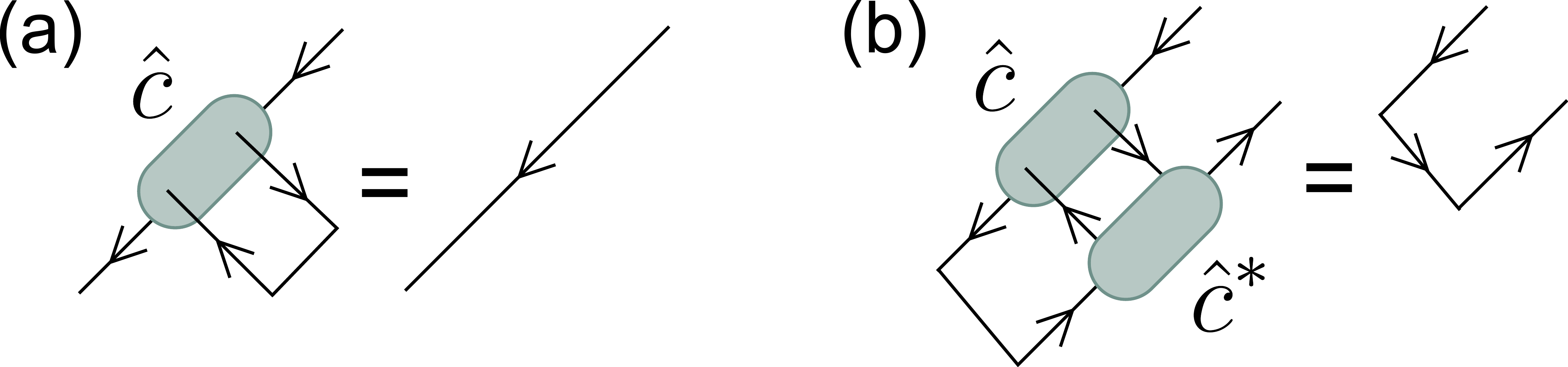}
\caption{\label{fig:copyTensor} Useful equalities satisfied by the $\mathcal{G}$-symmetric copy tensor $\hat{c}$. (a) Contraction of the identity on the open indices of $\hat{c}$ results in an identity. (b) Tensor $\hat{c}$ is an isometry.}
\end{figure}
%%%%%%%%%%%%%%%%%%%%%%%%%%%%%%%%%%%%%%%%%%%%%%%%%%%%%%%%%%%%%%%%%%%%%%%%%%%%%%%%%%%%%%%%%%%%%%%

\subsection{Some properties of the bulk state}\label{ssec:someprops}
The boundary state $\mypsi$ is recovered from a bulk state by projecting every pair of sites located on a bond to the state $\ket{+} \in (\mathbb{V} \otimes \mathbb{V})$ defined as
\begin{equation}
\ket{+} \equiv \sum_{a,t_a,m_a}  \ket{a,t_a,m_a} \otimes \ket{a^*,t_{a^*},m_{a^*}}. \nonumber
\end{equation}
%(Note that state $\ket{+}$ is invariant under the action of $\myg$, namely $\ket{+} = (\hat{V}_g \otimes \hat{V}^\dagger_g) \ket{+}$ for all $g \in \mathcal{G}$.)
State $\ket{+}$ is isomorphic to the identity matrix after the identification $\ket{a^*,t_{a^*},m_{a^*}} \leftrightarrow \bra{a,t_a,m_a}$. Thus, applying the projector $\hat{P}_k \equiv \ket{+}\bra{+}$ on the two bulk sites located on bond $k$ is equivalent to contracting the identity $\ket{+}$ with the copy tensor located on the bond. This contraction results in the identity, as depicted in \fref{fig:copyTensor}(a). Thus, the action of the projector $\hat{P}_k$ eliminates the copy tensor located on bond $k$ of the lifted MERA. By applying the projector on all the bonds of the lifted MERA, all the copy tensors are eliminated and we recover the MERA, and thus the boundary state $\mypsi$.

It is readily checked that the $\mathcal{G}$-symmetric copy tensor is an isometry satisfying the equality depicted in \fref{fig:copyTensor}(b). This, along with the fact that the MERA tensors $\hat{u}$ and $\hat{w}$ are isometries, ensures that a bulk state is normalized (see Ref.~\onlinecite{TNC}, Appendix A), and also exhibits the bulk features of the simpler lifted MERA described in Ref.~\onlinecite{TNC}, namely: (i) the presence of holographic screens, (ii) a simple dictionary that translates boundary correlators to expectation values of extended bulk operators, and (iii) a causal cone structure that can be exploited to compute bulk expectation values efficiently. These properties essentially rely on the fact that tensors $\hat{u}$ and $\hat{w}$ are isometries.

%0000000000000000000000000000000000000000000000000000000000000000000000000000000000000000000000000000
\section{Bulk entanglement}
\label{sec:bulkentanglement}

Given a subsystem of the bulk lattice $\mathcal{M}$, we define its perimeter and area as the number of sites that are located at the boundary and inside the subsystem respectively. For a generic state belonging to the lattice $\mathcal{M}$, subsystem entanglement entropy is expected to scale as the subsystem's area. In contrast, the subsystem entanglement entropy in a bulk state scales at most as the perimeter of the subsystem, see Appendix \ref{app:areaLaw}. Such an entanglement scaling is commonly exhibited by ground states of local Hamiltonians in condensed matter physics, where it is often called `area law entanglement' \cite{AreaLaw}. In fact, given a lifted MERA, which represents a bulk state $\myphi$, one can construct a local, gauge-invariant bulk Hamiltonian whose ground state is $\myphi$, as described in Appendix \ref{app:parentHam}.

In the remainder of this section we consider bulk states dual to 1D critical ground states, and explore any potential dependence of the bulk entanglement on the central charge of the CFT that describes the critical system in the continuum. We are motivated by the fact that in the AdS/CFT correspondence, the leading order of quantum fluctuations in the bulk is $O(1/c)$ where $c \gg 1$ is the central charge of the CFT \cite{quantumFlucBulk}. 

However, in order to compare bulk properties corresponding to different critical boundary states one has to address an ambiguity that arises from the fact that the MERA representation of a 1D ground state is not unique as discuss in the next section.

\subsection{Many bulk states dual to a ground state}\label{ssec:onetomany}
Given a MERA representation of a ground state, one can obtain another equivalent MERA representation of the state by inserting a resolution of identity $\hat{M}_k\hat{M}_k^{-1}$ on bond $k$, and multiplying the matrices $\hat{M}_k$ and $\hat{M}_k^{-1}$ with the two tensors that are connected by the bond respectively. The two MERAs are an equivalent representation of the ground state, since the expectation value of any observable is the same in both the representations. (Obtaining an expectation value from the MERA involves contracting all the bond indices, and $\hat{M}_k$ is multiplied with $\hat{M}_k^{-1}$ in the process.)

Clearly, inserting the copy tensor, defined according to \eref{eq:copycomponents}, selects out a particular MERA representation of the ground state---the one expressed in a bond basis in which the degeneracy tensors $\{\hat{c}^{\mbox{\tiny deg}}_a\}$ have these components. On the other hand, the degeneracy copy tensors $\{\hat{c}^{\mbox{\tiny deg}}_a\}$ `commute' only with diagonal matrices. Namely, a contraction of $\hat{c}^{\mbox{\tiny deg}}_a$ with a diagonal matrix on any index is equal to a contraction of the tensor with the same diagonal matrix on any other index. This implies that the  bulk states obtained by lifting different MERA representations of the same ground state are not generally related to each other by one-site unitary transformations on the bulk lattice, and therefore they have different entanglement. Thus, our bulk construction generally relates a given ground state to a set of bulk states with different entanglement.

However, in this paper, we restrict attention to MERA representations that are made of $\mathcal{G}$-\textit{symmetric} and \textit{isometric} tensors. While $\mathcal{G}$-symmetric tensors ensure that the bulk state---obtained by lifting the MERA by inserting copies of the $\mathcal{G}$-symmetric copy tensor---has a local gauge symmetry $\mathcal{G}$ (as described in Sec.~\ref{ssec:gaugesym}), the choice of isometric tensors leads to the desirable bulk features listed $(i)$-$(iii)$ in Sec.~\ref{ssec:someprops}.

To this end, we restrict $\{\hat{M}_k: \mathbb{V} \rightarrow \mathbb{V}\}_k$ to \textit{unitary} matrices that \textit{commute} with the symmetry, namely, $[\hat{M}_k,\hat{V}_g] = 0$ for all $g \in \mathcal{G}$. Since $\mathbb{V}$ decomposes as \eref{eq:symDecompose}, Schur's lemma (a special case of the Wigner-Eckart decomposition) implies that matrix $\hat{M}_k$ decomposes as $\hat{M}_k = \bigoplus_a (\hat{M}_{k,a} \otimes \hat{I}_{\eta_{a}})$. Thus, the bond transformations are restricted to act as the identity $\hat{I}_{\eta_{a}}$ on the bonds of the spin networks, which also restricts the set of the dual bulk states. In particular, one can exploit this restriction on the bond transformations to \textit{partially fix} a basis on the total bond space $\mathbb{V}$, in the different MERA representations of $\mypsi$. Specifically, we fix the irrep basis $\{\ket{a,m_a}\}$ on the bonds of the spin networks, while a basis on the bonds of the degeneracy tensor networks corresponds to a choice of the bond transformations $\hat{M}_{k,a}$ (with respect to a given MERA representation).

Therefore, here we probe for any \textit{statistical} dependence of the bulk entanglement on the boundary central charge, by randomly sampling from the set of all allowed dual bulk states. Recall that we only consider bulk states that are obtained by lifting MERA tensor networks composed of $\mathcal{G}$-\textit{symmetric} and \textit{isometric} tensors. (This corresponds to restricting the intrinsic bond transformations $\{\hat{M}_k\}$ to unitary matrices that commute with the symmetry $\mathcal{G}$.)

\subsection{Critical spin chains}
To this end, we considered the ground states of the following 1D critical spin models: 
\begin{equation}\label{eq:models}
\begin{split}
\hat{H}^{\mbox{\small \textsc{ising}}} &= \sum_{i} \hat{\sigma}^i_x\hat{\sigma}^{i+1}_x + \hat{\sigma}^i_z,\\
\hat{H}^{\mbox{\small \textsc{bc}}} &= \sum_{i} -\hat{S}_x^i\hat{S}_x^{i+1} + \alpha (\hat{S}_x^i)^2 + \beta \hat{S}_z^i,\\
\hat{H}^{\mbox{\small \textsc{potts}}} &= -\sum_{i} \hat{P}^i(\hat{P}^{T})^{i+1} + (\hat{P}^{T})^{i}\hat{P}^{i+1} + \hat{M}^i,\\
\hat{H}^{\mbox{\small \textsc{xxz}}} &= \sum_{i} \hat{\sigma}^i_x\hat{\sigma}^{i+1}_x + \hat{\sigma}^i_y\hat{\sigma}^{i+1}_y + \Delta \hat{\sigma}^i_z\hat{\sigma}^{i+1}_z,
\end{split}
\end{equation}
where $i$ labels sites of a 1D infinite lattice on which the Hamiltonian acts, $\hat{\sigma}_x,\hat{\sigma}_y,\hat{\sigma}_z$ are Pauli matrices, the operator $\hat{S}_{\alpha}$ is the $\alpha$ component of the spin$-1$ representation of $\mathfrak{su}(2)$, and $\hat{P}$ and $\hat{M}$ are $3 \times 3$ Potts matrices:
\[
\hat{P}=\left(\begin{array}{ccc}0 & 1 & 0 \\0 & 0 & 1  \\1 & 0 & 0\end{array}\right);\quad \hat{M}=\left(\begin{array}{ccc}2 & 0 & 0 \\0 & -1& 0 \\0 & 0 & -1\end{array}\right)
\]
The Blume-Capel model is critical for $\alpha=0.910207, \beta=0.415685$, and the XXZ model is critical for $-1 < \Delta \leq 1$.
The central charges and total symmetry groups of these models are listed in Table \ref{table:modelInfo}. 

%%%%%%%%%%%%%%%%%%%%%%%%%%%%%%%%%%%%%%%%%%%%%%%%%%%%%%%%%%%%%%%%%%%%%%%%%%%%%%%%%%%%%%%%%%%%%%%% slope reduction by 1/3
%\begin{tabular}[c]{@{}c@{}}\textsc{central charge}\\ \textsc{}\end{tabular}
\begin{table}[b]
\centering
\caption{The central charge and the total symmetry group of the critical lattice models listed in \eref{eq:models}.}
\begin{tabular}{|l|c|c|}
\hline
\multicolumn{1}{|c|}{\textsc{model}} & \multicolumn{1}{c|}{\textsc{central charge}} & \begin{tabular}[c]{@{}c@{}}\textsc{total}\\ \textsc{symmetry}\end{tabular} \\ \hline
Ising                                                                    & $\sfrac{1}{2}$                                                          & $Z_2$                                                         \\ \hline
Blume-Capel                                                     & $\sfrac{7}{10}$                                                    & $Z_2$
 \\ \hline
3-state Potts                                                             & $\sfrac{8}{10}$                                                       & $Z_3$
 \\ \hline

XXZ, $\Delta \neq 1$                                                   &  1                                                          & $U(1)$                                                          \\ \hline
XXZ, $\Delta=1$                                                                        & 1                                                        & $SU(2)$ 
\\ \hline
\end{tabular}\label{table:modelInfo}
\end{table}
%%%%%%%%%%%%%%%%%%%%%%%%%%%%%%%%%%%%%%%%%%%%%%%%%%%%%%%%%%%%%%%%%%%%%%%%%%%%%%%%%%%%%%%%%%%%%%%%

We determined a symmetry-protected MERA representation of the ground state of each of these models using the variational energy minimization algorithm for the scale-invariant MERA de-
scribed in Ref.~\onlinecite{MERACFT}, adapted to the presence of symmetries \cite{Singh,SinghU1}. We considered only Abelian symmetries here, which appear either as the total symmetry or as subgroup symmetry. Specifically, we obtained a $Z_2$-symmetric MERA representation for the ground state of the Ising model and the Blume-Capel model, and a $Z_3$-symmetric MERA representation of the ground state of the Potts model.

For the XXZ model, we obtained both a $Z_2$-symmetric MERA representation of the ground states for $\Delta \in \{0,0.71,0.81,0.87,1\}$ (corresponding to a $Z_2$ subgroup symmetry), and also an $U(1)$-symmetric MERA representation  for $\Delta \in \{0,1\}$. The $U(1)$ symmetry of the XXZ models with $\Delta \in \{0,1\}$ corresponds to the total symmetry for $\Delta=0$ and a subgroup symmetry for $\Delta=1$.

%%%%%%%%%%%%%%%%%%%%5
\begin{table}[t]
\centering
\caption{The representation of the symmetry on a each lattice site for the various 1D quantum lattice models listed in \eref{eq:models}. We use a compact notation $a(d_a)$ to denote an irrep $a$ and its degeneracy $d_a$ that appears in the irrep decomposition, \eref{eq:onesite}, of the Hilbert space of one site of the lattice. The two irreps of $Z_2$ are labelled by $0$ and $1$ respectively. The three irreps of $Z_3$ are labelled by $0,1$ and $2$ respectively. We label the two irreps of $U(1)$ that appear on each site of the XXZ model by $-1$ and $1$. For example, for the Blume-Capel model $0(2) \oplus 1(1)$ denotes that each site of the lattice decomposes as the direct sum of two copies of $Z_2$ irrep $0$ and one copy of  $Z_2$ irrep $1$. }
\begin{tabular}{ |l|l|l| }
\hline
\textsc{model} & \textsc{symmetry} & \textsc{site representation} \\ \hline
Ising & $Z_2$ & $0(1) \oplus 1(1)$ \\ \hline
Blume-Capel & $Z_2$ & $0(2) \oplus 1(1)$ \\ \hline
Potts & $Z_3$ & $0(1) \oplus 1(1) \oplus 2(1)$ \\ \hline
XXZ & $Z_2$ & $0(1) \oplus 1(1)$ \\ \hline
XXZ, $\Delta=0,1$ & $U(1)$ & $-1(1) \oplus 1(1)$ \\ \hline
\end{tabular}\label{table:repsite}
\end{table}
%%%%%%%%%%%%%%%%%%%%%%%%%%%%%%%

%%%%%%%%%%%%%%%%%%%%5
\begin{table}[t]
\centering
\caption{The symmetry representation that we fixed on the MERA bonds in the ground state simulation of the 1D quantum lattice models listed in \eref{eq:models}. Since the bonds of the MERA are associated with coarse-grained sites, the bond representation is obtained by fusing and truncating the symmetry representations that appear on multiple sites of the 1D lattice. (The symmetry representation on each site of the lattice is listed in Table \ref{table:repsite}.) The total bond dimension (namely, the dimension of the total bond representation) is equal to 12 for all the simulations. }
\begin{tabular}{ |l|l|l| }
\hline
\textsc{model} & \textsc{symmetry} & \textsc{bond representation} \\ \hline
Ising & $Z_2$ & $0(6) \oplus 1(6)$ \\ \hline
Blume-Capel & $Z_2$ & $0(6) \oplus 1(6)$ \\ \hline
Potts & $Z_3$ & $0(4) \oplus 1(4) \oplus 2(4)$ \\ \hline
XXZ & $Z_2$ & $0(6) \oplus 1(6)$ \\ \hline
XXZ, $\Delta=0,1$ & $U(1)$ & $-3(2) \oplus -1(4) \oplus 1(4)  \oplus 3(2)$ \\ \hline
\end{tabular}\label{table:repbond}
\end{table}
%%%%%%%%%%%%%%%%%%%%%%%%%%%%%%%

%We kept the same bond dimension $\chi = 12$ in all the ground state simulations, with the choice of charges and degeneracies shown in 
The representation of the symmetry on each site of the lattice for the various models is listed in Table \ref{table:repsite}. Table \ref{table:repbond} lists the symmetry representation that we assigned to the MERA bonds for the ground state simulations. We tried a few different charge and degeneracy combinations and the choices listed in the table \ref{table:repbond} resulted in the smallest error in the ground state energy density as determined from the resulting MERA.

The error in the estimated ground state energy density for the Ising model was $O(10^{-8})$ and the relative error in the estimated central charge was $0.6\%$. For the remaining models, the error in the estimated ground state energy density was at most $O(10^{-4})$, and the relative error in the estimated central charge was at most $3\%$. For all the models, the relative error in the estimated smallest six scaling dimensions was between $0.5\%$ to $6\%$.

The models listed in \eref{eq:models} are not scale-invariant, but flow to a scale-invariant fixed point after possibly several RG (entanglement renormalization) steps. We discarded the non-scale invariant part of the MERA before lifting it to a bulk state. (That is, we considered the renormalized scale-invariant ground state of each model.)

%%%%%%%%%%%%%%%%%%%%%%%%%%%%%%%%%%%%%%%%%%%%%%%%%%%%%%%%%%%%%%%%%%%%%%%%%%%%%%%%%%%%%%%%%%%%%%%
\begin{figure}[t]
\includegraphics[width=\columnwidth]{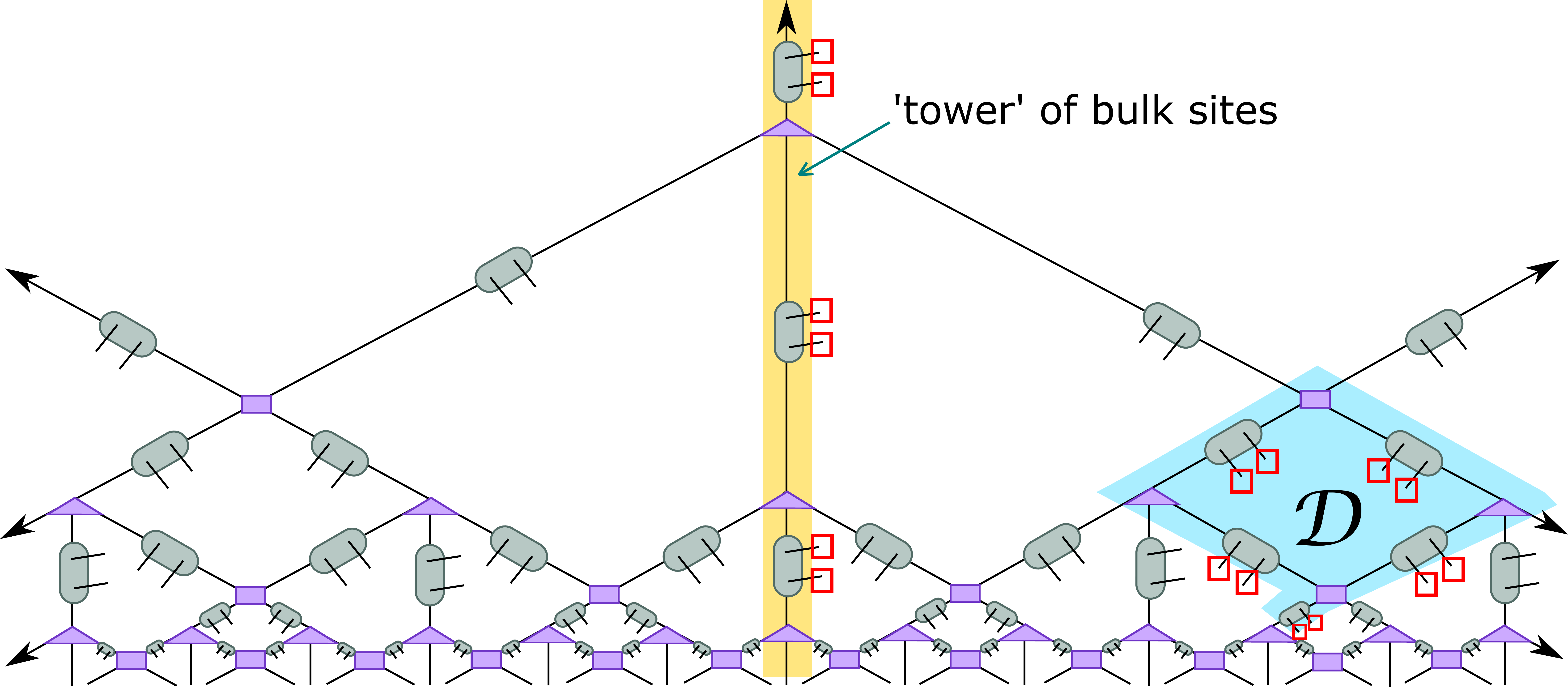}
\caption{\label{fig:dongle} The bulk sites (red squares) located in the region highlighted yellow were considered to obtain the plots shown in \fref{fig:TIMDistribution} and \fref{fig:BoxPlots}. These sites are located along an infinitely long tower of the $\hat{w}$ tensors. The bulk sites located in the region highlighted blue were considered to obtain the entanglement negativities listed in Table \ref{table:negativity}. These consist of sites located around a loop of tensors and two sites located at the bottom of the loop.
%This is the smallest bulk region that exhibits non-zero entanglement negativity for any bulk state.
}
\end{figure}
%%%%%%%%%%%%%%%%%%%%%%%%%%%%%%%%%%%%%%%%%%%%%%%%%%%%%%%%%%%%%%%%%%%%%%%%%%%%%%%%%%%%%%%%%%%%%%%

%%%%%%%%%%%%%%%%%%%%%%%%%%%%%%%%%%%%%%%%%%%%%%%%%%%%%%%%%%%%%%%%%%%%%
\begin{figure}[b]
\includegraphics[width=\columnwidth]{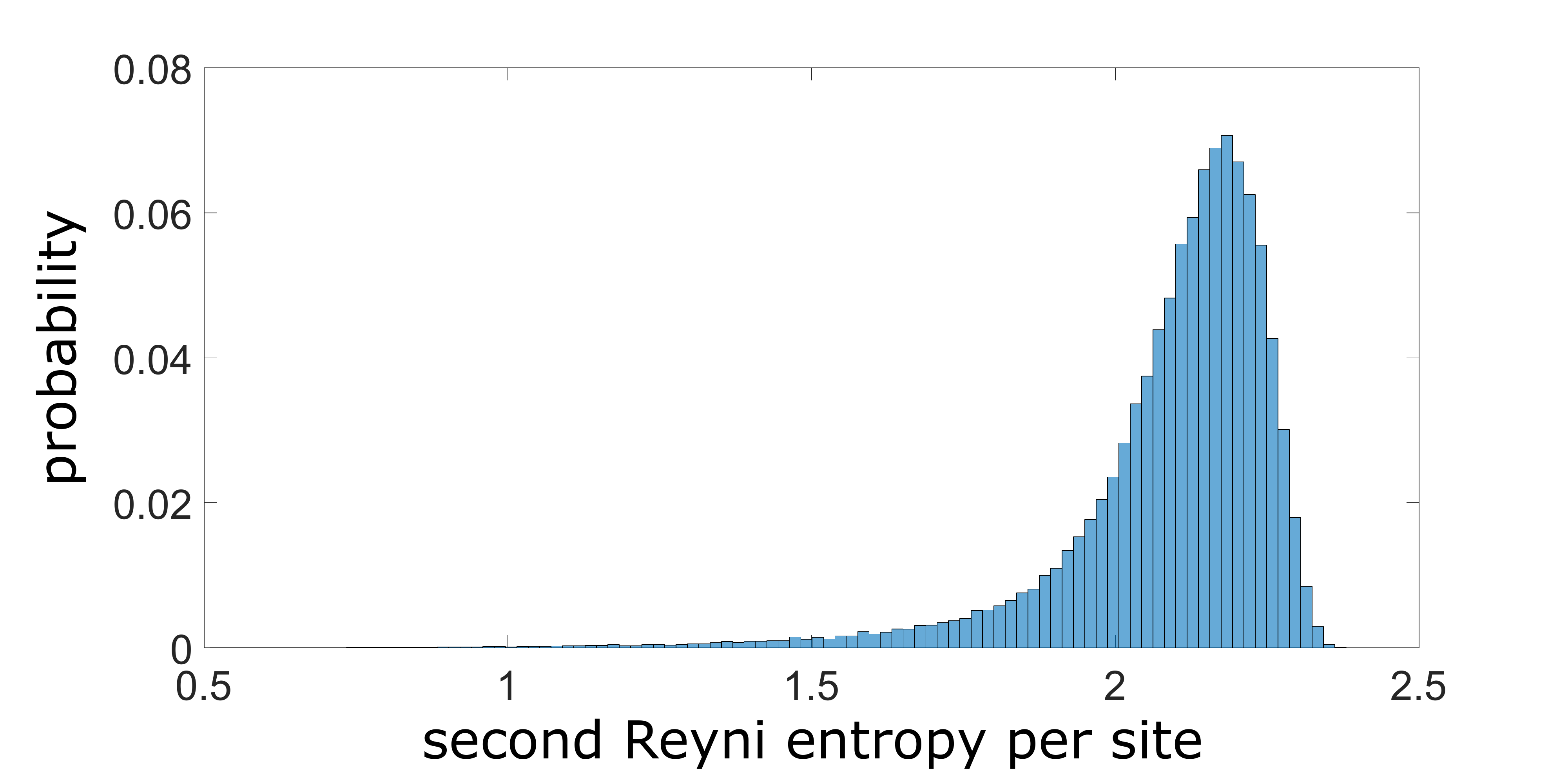}
\caption{\label{fig:TIMDistribution} A probability distribution of the Renyi entanglement entropy $R^{\mbox{\tiny tower}}$ per site [\eref{eq:rhogeo}], computed from randomly sampled bulk states dual to the ground state of the critical Ising model. We sampled $10^{5}$ bulk states and sorted the corresponding entropy densities into 100 equally spaced bins.}
\end{figure}
%%%%%%%%%%%%%%%%%%%%%%%%%%%%%%%%%%%%%%%%%%%%%%%%%%%%%%%%%%%%%%%%%%%%%

\subsection{Bulk entanglement vs boundary central charge}\label{ssec:entanglementPlot}
Before proceeding to our results, we remark that defining entanglement entropy in gauge-invariant states is subtle since a gauge-invariant Hilbert space does not usually have a tensor product structure. A possible approach, one that we have followed here, is to embed the Hilbert space into a larger tensor product space---the tensor product of the Hilbert spaces on each of the links of a lattice gauge theory. See, for example, a recent work presented in Ref.~\onlinecite{GaugeEntanglement} and references contained therein.

For the ground state of each of the models listed in \eref{eq:models}, we randomly selected $10^{5}$ dual bulk states (restricting the corresponding bond transformations $\{\hat{M}_k\}$ to unitary matrices that commute with the respective symmetry), and computed the second Renyi entanglement entropy $R^{\mbox{\tiny tower}}$ per site,
\begin{equation}\label{eq:rhogeo}
R^{\mbox{\tiny tower}} \equiv -\mbox{log}_2 (\mbox{Tr}(\hat{\rho}^{\mbox{\tiny tower}})^2).
\end{equation}
Here $\hat{\rho}^{\mbox{\tiny tower}}$ is the reduced density matrix of all the bulk sites located along the infinitely long tower of $\hat{w}$ tensors (highlighted yellow in \fref{fig:dongle}). We partitioned the Renyi entanglement entropy density values in to 100 equally spaced bins.

Figure \ref{fig:TIMDistribution} shows the probability distribution of the Renyi entanglement entropy $R^{\mbox{\tiny tower}}$ per site for the critical Ising model, which illustrates that the different bulk states indeed have different entanglement.

%%%%%%%%%%%%%%%%%%%%%%%%%%%%%%%%%%%%%%%%%%%%%%%%%%%%%%%%%%%%%%%%%%%%%
\begin{figure}[t]
\includegraphics[width=7cm]{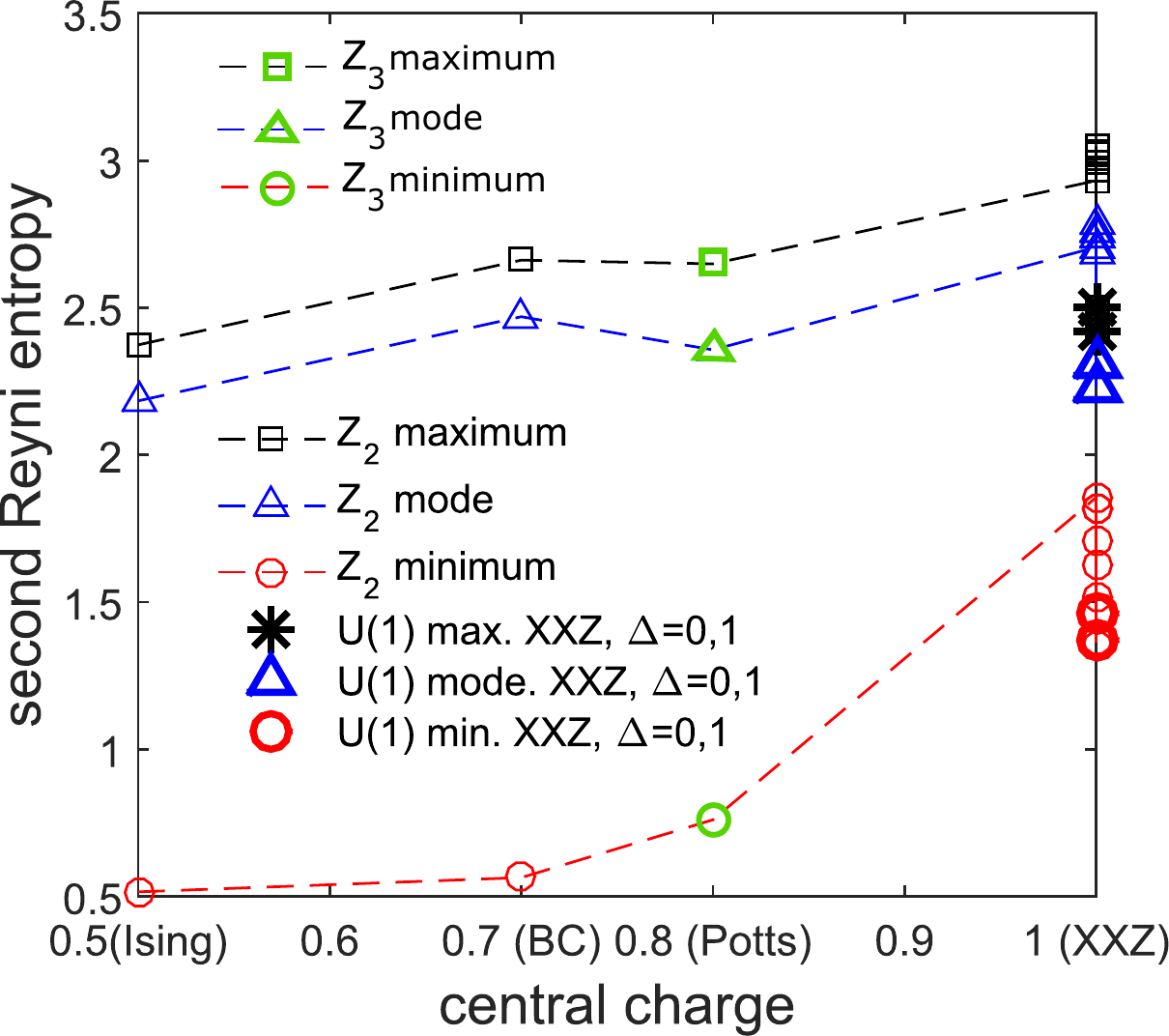}
\caption{\label{fig:BoxPlots} The maximum (square, star), mode (triangle), and minimum (circle) of the probability distribution of the Renyi entanglement entropy $R^{\mbox{\tiny tower}}$ per site [\eref{eq:rhogeo}] per site, obtained from randomly sampled $10^{5}$ bulk states, dual to the ground state of each of the critical models listed along the $x$-axis. The various $Z_2$ data for XXZ models correspond to $\Delta \in \{0,0.71,0.81,0.87,1\}$. Explanation in Sec.~\ref{ssec:entanglementPlot}.}
%The plot indicates a statistical trend that the Renyi entropy density generally increases with the boundary central charge. The plot also suggests that a MERA representation for larger subgroup symmetry may correlate with a decrease in the entropy.}
\end{figure}
%%%%%%%%%%%%%%%%%%%%%%%%%%%%%%%%%%%%%%%%%%%%%%%%%%%%%%%%%%%%%%%%%%%%%

In \fref{fig:BoxPlots} we plot the maximum, minimum and the mode of the probability distributions of $R^{\mbox{\tiny tower}}$ per site for all the critical models. The plot indicates a statistical trend that the Renyi entropy density generally increases with increase in the boundary central charge. Note also the clustering of data for different XXZ models, which have the same central charge. From these results it appears that the bulk entanglement entropy $R^{\mbox{\tiny tower}}$ depends predominantly on the central charge, as compared to  other microscopic details of these models.

The plot in \fref{fig:BoxPlots} also suggests that a MERA representation based on a larger subgroup symmetry may correlate with a decrease in the Renyi entropy $R^{\mbox{\tiny tower}}$. Specifically, for ground states of the two XXZ models with $\Delta=0,1$ the maximum, mode and minimum Renyi entropies obtained from the $U(1)$-symmetric MERA representation were found to be smaller than those obtained from the $Z_2$-symmetric MERA representation. On the other hand, the two representations gave approximately equal estimates for the ground state energies, central charges, and few lowest scaling dimensions.

Note that a $Z_2$-symmetric and a $U(1)$-symmetric MERA representation of a given ground state are expected to correspond to two bulk states with different entanglement respectively. This is because a $U(1)$-symmetric MERA representation can be converted to a $Z_2$-symmetric MERA representation by applying bond transformations to change the bond basis from a $U(1)$ irrep basis to the subgroup $Z_2 $ basis listed in Table \ref{table:repbond}, which likely alter the bulk entanglement. However, we do not know how to account for the decrease in entropy when the larger symmetry was considered here, and whether this behaviour is more general than illustrated by these results.

\subsection{Entanglement between gauge and degeneracy degrees of freedom}\label{ssec:emergent}
Finally, we probed the bulk entanglement between the gauge and degeneracy degrees of freedom for the case of $Z_2$ and $Z_3$ symmetry. We considered a small region of the bulk lattice and obtained a reduced density matrix by tracing out all degrees of freedom outside the region, \textit{and also the gauge degrees of freedom inside the region}. This was achieved by using the spin network decomposition of the bulk state, which exposes separate open indices in the lifted MERA corresponding to the gauge and non-gauge degrees of freedom respectively. In order to trace out the gauge degrees of freedom in a region, one also contracts the open indices of the spin networks that are located with the region but not the corresponding degeneracy indices.

%%%%%%%%%%%%%%%%%%%%%%%%%%%%%%%%%%%%%%%%%%%%%%%%%%%%%%%%%%%%%%%%%%%%%%%%%%%%%%%%%%%%%%%%%%%%%%%% slope reduction by 1/3
\begin{table}[b]
\centering
\caption{The entanglement negativity $n(\hat{\rho}^{\scriptscriptstyle [\mathcal{D}]})$, \eref{eq:negativity}, obtained from two different bulk states, dual to the ground state of each of the critical models listed in \eref{eq:models}. Here we used a $Z_3$-symmetric MERA for the Potts model and a $Z_2$-symmetric MERA for the remaining models.}
\begin{tabular}{|l|c|c|}
\hline
\multicolumn{1}{|c|}{\textsc{model}} & \begin{tabular}[c]{@{}c@{}}\textsc{Bulk state 1}\end{tabular} & \begin{tabular}[c]{@{}c@{}}\textsc{Bulk state 2}\end{tabular} \\ \hline
Ising                                                                    & 0.01953                                                          & 0.08136                                                         \\ \hline
Blume-Capel                                                           & 0.09975                                                        & 0.92443                                                 \\ \hline
3-state Potts                                                             & 0.08258                                                       & 0.37165                                                 \\ \hline
XXZ, $\theta=1$                                                   &  0.04061                                                          & 0.61028                                                          \\ \hline
XXZ, $\theta=0$                                                                        & 0.05483                                                          & 0.24789                                                          \\ \hline
\end{tabular}\label{table:negativity}
\end{table}
%%%%%%%%%%%%%%%%%%%%%%%%%%%%%%%%%%%%%%%%%%%%%%%%%%%%%%%%%%%%%%%%%%%%%%%%%%%%%%%%%%%%%%%%%%%%%%%%

The smallest region for which we found non-zero \textit{entanglement negativity}, a measure of quantum entanglement, is depicted as region $\mathcal{D}$ in \fref{fig:dongle}. Let $\hat{\rho}^{\scriptscriptstyle [\mathcal{D}]}$ denote the reduced density matrix of region $\mathcal{D}$ by tracing out all bulk sites outside $\mathcal{D}$, and also the gauge degrees of freedom inside $\mathcal{D}$. We computed the entanglement negativity $n(\hat{\rho}^{\scriptscriptstyle [\mathcal{D}]})$ given by
\begin{equation}\label{eq:negativity}
n(\hat{\rho}^{\scriptscriptstyle [\mathcal{D}]}) = \sum_i (|\lambda_i| - \lambda_i)/2, 
\end{equation}
where $\lambda_i$ are the eigenvalues of the matrix obtained by taking the partial transpose of $\hat{\rho}^{\scriptscriptstyle [\mathcal{D}]}$ with respect to some of the sites in $\mathcal{D}$. (A non-zero value of the negativity indicates that the state has quantum entanglement.) 
We selected two different bulk states dual to the ground state of each critical model listed in \eref{eq:models}, and computed the value of $n(\hat{\rho}^{\scriptscriptstyle [\mathcal{D}]})$ for both these bulk states. These values are listed in Table \ref{table:negativity}. The fact that this entanglement negativity is positive for these models indicate that (at least) these dual bulk states have quantum entanglement between the gauge and degeneracy degrees of freedom.

%%%%%%%%%%%%%%%%%%%%%%%%%%%%%%%%%%%%%%%%%%%%%%%%%%%%%%%%%%%%%%%%%%%%%
\begin{figure}[t]
\includegraphics[width=7cm]{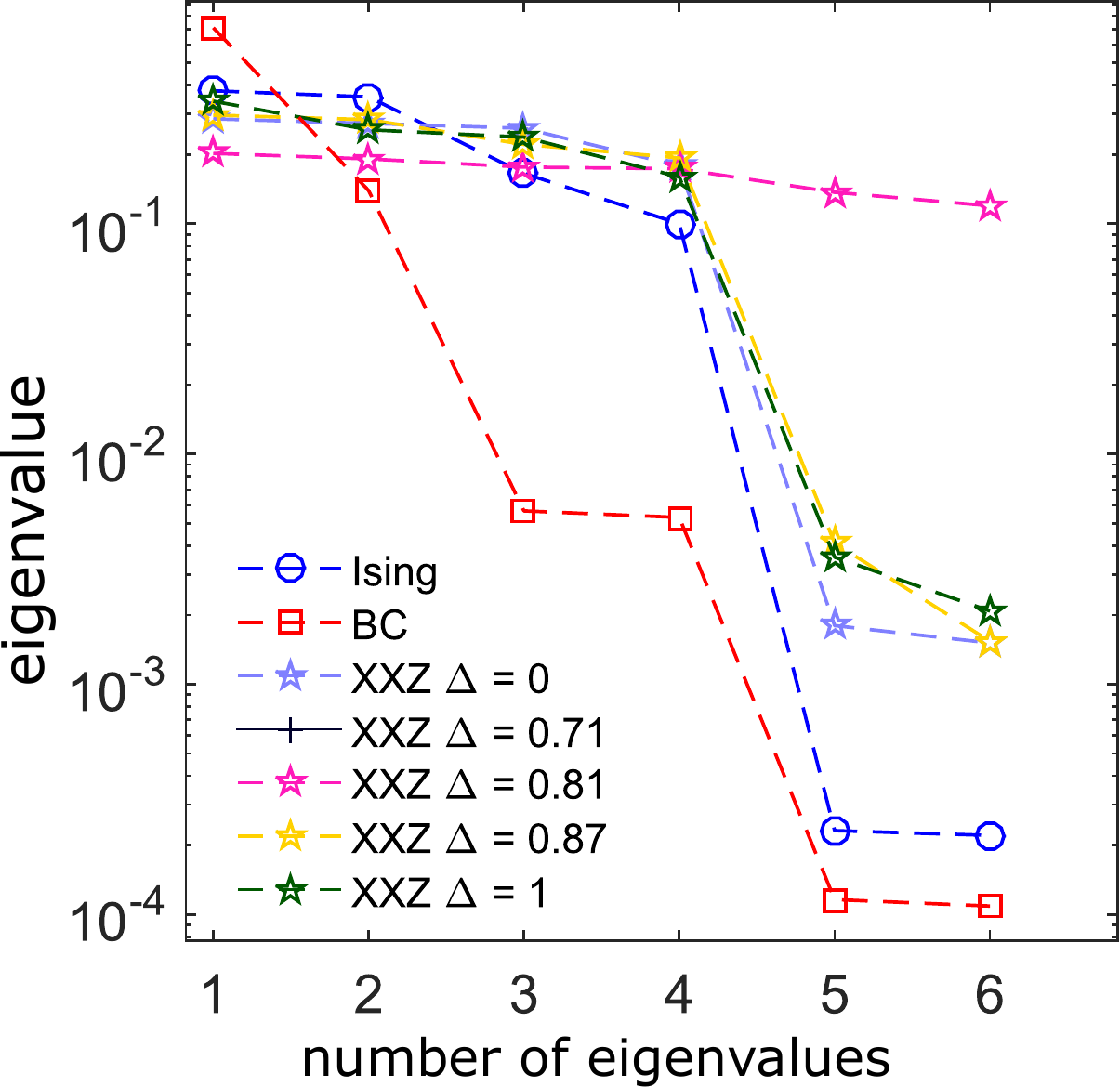}
\caption{\label{fig:degspec} The spectrum of a reduced density matrix obtained from a randomly selected bulk state, dual to the ground state of each of the critical models listed in \eref{eq:models}. The reduced density matrix corresponds to one bulk site, obtained by tracing out all remaining bulk sites and also tracing out the gauge degrees of freedom of that site.}
\end{figure}
%%%%%%%%%%%%%%%%%%%%%%%%%%%%%%%%%%%%%%%%%%%%%%%%%%%%%%%%%%%%%%%%%%%%%

The plot in \fref{fig:degspec} shows the spectrum of the reduced density matrix $\hat{\rho}_k^{\tiny \mbox{deg}}$, obtained by tracing out all bulk sites except the bulk site $k$ and also tracing out the gauge degrees of freedom on the site $k$. (That is, $\hat{\rho}_k^{\tiny \mbox{deg}}$ has support only on the non-gauge sector of site $k$.) Notice the appearance of approximate degeneracies in the spectrum. (We sampled a few different bulk states, this plot is illustrative of the typical degeneracies that we observed.) 

One possible way to account for these degeneracies is the emergence of a non-Abelian symmetry in the non-gauge (`gravitational') sector of the bulk. For instance, if the bulk state has an (emergent) non-Abelian on-site symmetry, say which acts only on the non-gauge sector of the Hilbert space, then the reduced density matrix $\hat{\rho}_k^{\tiny \mbox{deg}}$ must commute with this symmetry. Consequently, by applying Schur's lemma, $\hat{\rho}_k^{\tiny \mbox{deg}}$ must decompose as 
\begin{equation}
\hat{\rho}^{\mbox{\tiny deg}}_{k} = \bigoplus_j (\hat{\rho}^{\mbox{\tiny deg}}_{k,j} \otimes \hat{I}_{\eta_j}).
\end{equation}
Here $j$ is an irrep of the emergent symmetry, $\hat{\rho}^{\mbox{\tiny deg}}_{k,j}$ is a density matrix that acts on the degeneracy space of charge $j$, and $\hat{I}_{\eta_j}$ is the $\eta_j \times \eta_j$ identity matrix that acts on the irrep $j$. The spectrum of $\hat{\rho}_k^{\tiny \mbox{deg}}$ is clearly degenerate, in accordance with this decomposition, specifically, the degeneracy of an eigenvalue of $\hat{\rho}^{\mbox{\tiny deg}}_{k,j}$ is at least $\eta_j$.

In the scenario of an emergent symmetry, the degeneracies in the spectrum of $\hat{\rho}^{\mbox{\tiny deg}}_{k}$ can be used to infer a possible set of emergent symmetry charges, which can be used to decorate the bonds of the degeneracy tensor networks that appear in \fref{fig:spinnetwork} (analogous to how the spin networks are decorated with the symmetry charges). Broadly speaking, in this case, it may be possible to further decompose the degeneracy tensor networks in terms of spin networks composed of intertwiners of the emergent symmetry, thus refining the bulk construction presented in this paper. We leave further exploration of any emergent bulk symmetries for future work.

%00000000000000000000000000000000000000000000000000000000000000000000000000000000000000000000000000
\section{Summary and Outlook}\label{sec:outlook}

In this paper, we described a toy model for constructing a holographic description of a 1D quantum lattice system, equipped with the action of a local Hamiltonian that has a global onsite symmetry $\mathcal{G}$. Specifically, we lifted a MERA representation of the ground state, which also has the global symmetry, to a tensor network representation of a quantum state of a 2D lattice on which the symmetry appears gauged.
This was achieved by embedding the MERA in a 2D manifold, and inserting 4-index tensors on the bonds of the tensor network. The 1D ground state and the dual 2D quantum state are seen to live on the boundary and in the bulk of the manifold respectively. In order to manifest a gauge symmetry in the bulk, it was essential to use $\mathcal{G}$-symmetric tensors, which compose the MERA representation and generate a symmetry protected RG flow, and require that the copy tensors, which were used to lift the MERA, fulfill particular symmetry properties, those depicted in \fref{fig:copy}.

In this way, our toy model translates a 1D \textit{boundary} state with a global onsite symmetry to a 2D \textit{bulk} state in which the symmetry appears gauged. In the AdS/CFT correspondence, a global onsite symmetry at the boundary is also gauged in the bulk as a general rule of thumb. In light of the ongoing discussion between the MERA and holography, we take the view that any legitimate bulk description of the MERA must implement the holographic gauging of global boundary symmetries. In particular, making this demand may narrow the choices for the possible bulk degrees of freedom. As we have shown in this paper, the holographic gauging of boundary symmetries is very conveniently realized by associating bulk degrees of freedom with the bonds of the MERA, as opposed to its tensors as has been considered in some of the previous works \cite{ExactHolo,HoloCode,HoloRandom}, since it allows us to introduce gauge transformations as in lattice gauge theory.

%We also explored the entanglement of in the dual bulk states. For example, we illustrated a potential (statistical) dependence of the bulk entanglement on the boundary central charge.
We further showed how the bulk states decompose as a superposition of spin network states, which label a basis in the gauge-invariant sector of the bulk Hilbert space. Thus, our bulk construction brings together tensor network states and spin network states, as they appear in lattice gauge theories with gauge group $\mathcal{G}$.
%We remark on a few broad differences between spin network and tensor network descriptions of many-body states. A spin network state is described by means of a spin network---a tensor network composed of intertwiners of the symmetry---and constitutes a basis in the gauge-invariant sector of the Hilbert space. The degrees of freedom of a spin network state are associated with the bond indices of the spin network. And in lattice gauge theories spin network states label a basis in the gauge-invariant sector of the Hilbert space. On the other hand, the degrees of freedom of a tensor network state are associated with only the open indices of the tensor network.
%We remark that the bulk construction described in this paper may be generalized by also exposing and lifting the internal intertwining charges (that is, the $e_i$'s that appear in \fref{fig:spinnetwork} also appear as open indices in the lifted MERA), which allows to incorporate `gauge matter' in the model.
The spin network decomposition of the bulk state allows one to introduce meaningful gauge-invariant observables in the bulk, since it exposes quantum numbers in the bulk (within the gauge-invariant sector). It also allows us to explore further parallels with holography. For example, the decomposition reveals correlations between the gauge and the remaining (`gravitational') degrees of freedom, and is also instrumental to probe any emergent symmetries in the non-gauge (`gravitational') degrees of freedom (since the spin network decomposition allows one to trace out the gauge degrees of freedom in the bulk). 

Spin networks also appear in various quantum gravity models where they label a gauge-invariant basis in the kinematic Hilbert space of the theory, for example, in loop quantum gravity \cite{LQGSpinNetworks}. Towards the completion of this work, we found recent papers which also explore connections between tensor network states and spin network states, specifically as they appear in loop quantum gravity \cite{LQGExactHolo}, and in the context of group field theory \cite{GroupFieldTensorNetwork}.

This work demonstrates a useful toy model for exploring basic features of holography using tensor networks.
Beyond holography, our formalism may be viewed as a general correspondence between a 1D ground state with a global symmetry $\mathcal{G}$ and a 2D many-body state with a local symmetry $\mathcal{G}$, which may also be useful in characterizing and relating together different types of quantum phases of matter as illustrated in Appendix~\ref{sec:holonomy}.

\textbf{Acknowledgements.}---This research was largely completed while SS was employed at the Center for Engineered Quantum Systems in Macquarie University. We thank Guifre Vidal, John Baez, Sundance Bilson-Thompson, Yichen Shi, Giandemenico Palumbo and Rob Pfeifer for stimulating discussions. SS acknowledges the hospitality of the Perimeter Institute for Theoretical Physics where a part of this work was presented. We acknowledge support from the ARC via the Centre of Excellence in Engineered Quantum Systems (EQuS), project number CE110001013 and from DP160102426.

%%%%%%%%%%%%%%%%%%%%%%%%5

\appendix

 %%%%%%%%%%%%%%%%%%%%%%%%%%%%%%%%%%%%%%%%%%%%%%%%%%%%%%%%%%%%%%%%%%%%%%%%%%%%%%%%%%%%%%%%%%%%%%%%%
\begin{figure}[t]
  \includegraphics[width=\columnwidth]{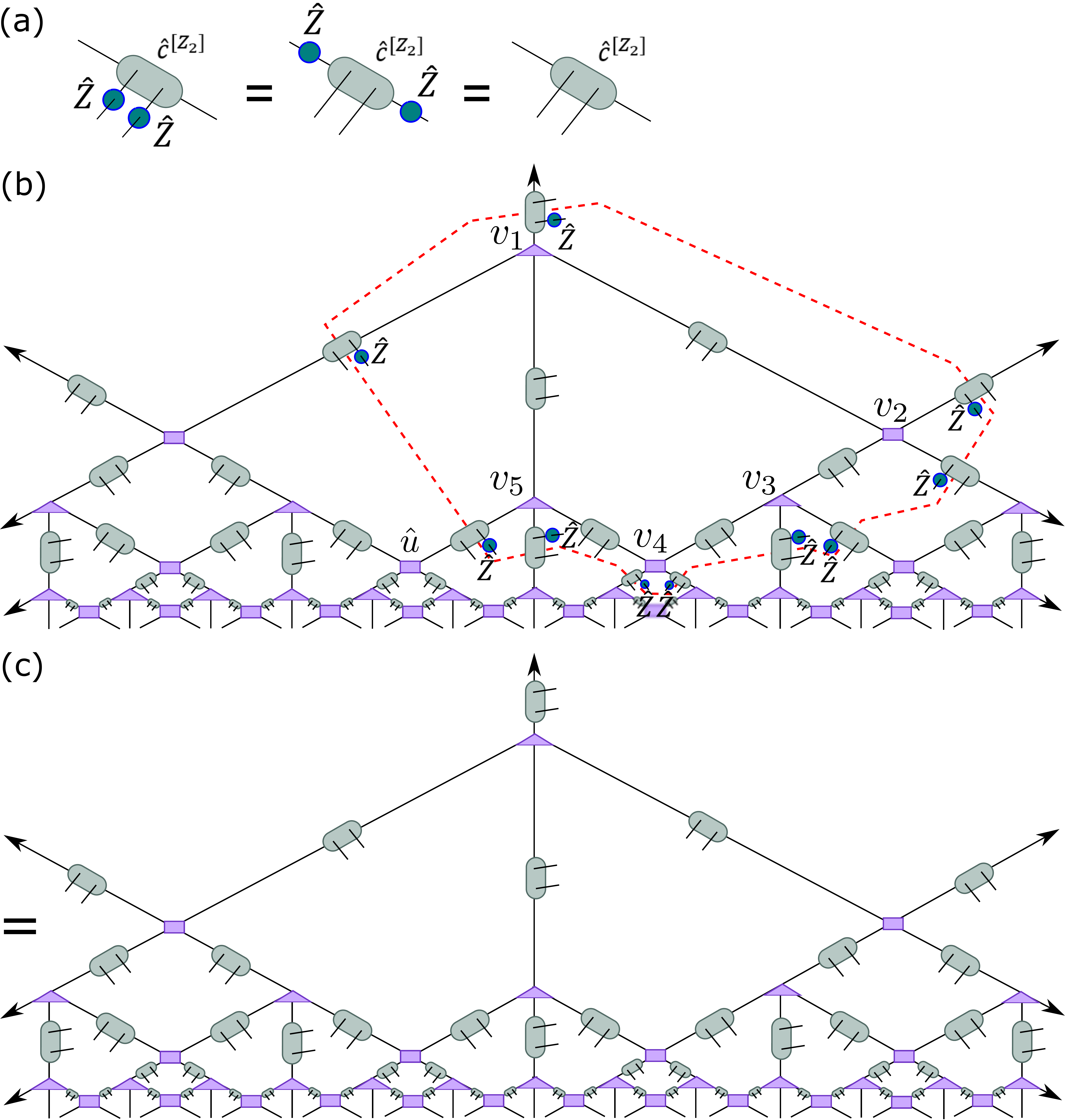}
\caption{\label{fig:zLoop} (a) The $Z_2$-symmetric copy tensor $\hat{c}^{\zz}$, \eref{eq:z2copy} is invariant under the action of $\hat{Z}$ on any two of its indices. (b,c) The lifted MERA $\mathcal{T}'$, and thus the bulk state it represents, is invariant under the action of the loop (dashed red contour) of $\hat{Z}$'s shown here. Namely, the contraction depicted in $(b)$ simply recovers the lifted MERA $(c)$.}
\end{figure}
%%%%%%%%%%%%%%%%%%%%%%%%%%%%%%%%%%%%%%%%%%%%%%%%%%%%%%%%%%%%%%%%%%%%%%%%%%%%%%%%%%%%%%%%%%%%%%%

%00000000000000000000000000000000000000000000000000000000000000000000000000000000000000000000000000000000000
\section{Bulk topological order and isometric tensors}\label{sec:holonomy}
In this appendix, we discuss the topological order of the bulk states obtained from the MERA as described in this paper, applied to the case of $Z_2$ symmetry. (The discussion can be readily generalized to $Z_n$ symmetry.) In particular, we illustrate an interesting interplay between bulk (Abelian) topological order and the choice of isometric tensors. More specifically, the bulk state obtained by lifting a MERA made of $Z_n$-symmetric and isometric tensors does not have a non-trivial $Z_n$ topological order.

\subsection{Example using $Z_2$ symmetry}
For the purpose of this section, we specialize the notation introduced in Sec.~\ref{sec:boundary} to the case of a $Z_2$ symmetry. Let $\mathcal{L}$ here denote an infinite 1D lattice, each site of which is described by vector space $\mathbb{V} \cong \mathbb{C}_2$ and is equipped with the action of the group $Z_2=\{\hat{I},\hat{Z}\}$. The group acts on the space $\mathbb{V}$ by means of the unitary representation
$\hat{I} = \smallmat{1}{0}{0}{1},~\hat{Z} = \smallmat{1}{0}{0}{-1}$.
Under the action of the symmetry, the space $\mathbb{V}$ decomposes as
\begin{equation}\label{eq:bondspace}
\mathbb{V} \cong \mathbb{V}_e \oplus \mathbb{V}_o,\nonumber
\end{equation}
where $\mathbb{V}_e$ and $\mathbb{V}_o$ are the two irreps of $Z_2$. We denote by $\ket{e \equiv 0}$ and $\ket{o \equiv 1}$ a basis in the one dimensional vector spaces $\mathbb{V}_e$ and $\mathbb{V}_o$ respectively.

Let $\mypsisym$ denote the (unnormalized) GHZ state belonging to the lattice $\mathcal{L}$,
\begin{equation}\label{eq:ghz1}
\mypsisym \equiv \ket{++\cdots} + \ket{--\cdots},
\end{equation}
where $\ket{\pm} = (\ket{e}\pm\ket{o})$ and e.g. $\ket{++\cdots} \equiv (\ket{+} \otimes \ket{+} \otimes \cdots)$. State $\mypsisym$ has a global $Z_2$ symmetry since $\mypsisym = (\hat{Z} \otimes \hat{Z} \otimes \ldots) \mypsisym$.

Consider a  $Z_2$-symmetric MERA representation $\mathcal{T}$ of $\mypsisym$ comprised of copies of two simple tensors
\begin{equation}
\ughz:\mathbb{V} \otimes \mathbb{V} \rightarrow \mathbb{V} \otimes \mathbb{V},~~~\wghz: \mathbb{V} \rightarrow \mathbb{V} \otimes \mathbb{V} \otimes \mathbb{V},
\end{equation}
which replace copies of the tensors $\hat{u}$ and $\hat{w}$ in \fref{fig:mera} respectively.
Tensor $\ughz$ is simply the identity, 
\begin{equation}
(\ughz)^{ij}_{kl} = \delta^{i}_{k}\delta^{j}_{l},~~~i,j,k,l \in \{e,f\},
\end{equation}
and the components of $\wghz$ are:
 \begin{equation}\label{eq:z2MERA}
    (\wghz)^{i}_{jkl} =
    \begin{cases}
      1, & \text{if } (i+j+k+l)\text{ mod } 2 =0 \\
      0, & \text{otherwise.}
    \end{cases}
  \end{equation}
Note that tensor $\wghz$ is an isometry satisfying
\begin{equation}
\sum_{ijk}(\wghz)^{i}_{jkl}(\wghz{}^*)_{i'}^{jkl} = \delta^i_{i'}.
\end{equation}
It is readily checked that tensor $\wghz$ is also $Z_2$-symmetric, since $\wghz = (\hat{Z}) \wghz  (\hat{Z}^\dagger \otimes \hat{Z}^\dagger \otimes  \hat{Z}^\dagger)$.

In order to verify that the MERA tensor network $\mathcal{T}$ indeed represents the state $\mypsisym$, \eref{eq:ghz1}, we need to to contract all the tensors of $\mathcal{T}$ to obtain the probability amplitudes $\hat{\Phi}_{i_1i_2\cdots}$, where $i_1i_2\cdots \in \{e,o\}$ denote the open indices of $\mathcal{T}$. The simple tensor network $\mathcal{T}$ can be contracted algebraically to obtain
 \begin{equation}\label{eq:ghz}
    \hat{\Phi}_{i_1i_2\cdots} = 
    \begin{cases}
      1, & \text{if }  \sum_{k} i_k \text{ is even}, \\
      0, & \text{otherwise.}
    \end{cases}
  \end{equation}
That is, $\mypsisym$ is an (unnormalized) equal superposition of kets $\ket{o_1o_2\cdots} \equiv \ket{o_1} \otimes \ket{o_2} \otimes \cdots$ labelled by bit strings $o_1o_2\cdots$ with even number of 1's. In the basis $\ket{\pm}$ on each site, this state is simply the GHZ state $\mypsisym$.

%00000000000000000000000000000000
%\subsection{Bulk state}
Let us lift the tensor network $\mathcal{T}$ by inserting the $Z_2$-symmetric copy tensor $\hat{c}^{\zz}$ on each bond of the tensor network. The copy tensor $\hat{c}^{\zz}$ is defined by specializing \eref{eq:bondtensor} and \eref{eq:copycomponents} to $Z_2$, namely, 
 \begin{equation}\label{eq:z2copy}
    (\hat{c}^{\zz})^{ij}_{kl} = 
    \begin{cases}
      1, & \text{if }  i=j=k=l, \\
      0, & \text{otherwise.}
    \end{cases}
  \end{equation}
where $i,j,k,l = \{e,o\}$. By construction, the bulk state $\myphisym$ represented by this lifted tensor network has a local $Z_2$ gauge symmetry, as described in Sec.~\ref{sec:bulk}.

%00000000000000
%\subsection{Topological order of the bulk state}\label{ssec:bulktopo}
We now ask whether the bulk state $\myphisym$ has $Z_2$ topological order? In order for the state $\myphisym$ to have a $Z_2$ topological order it must be invariant under the action of two non-commuting, deformable loop operators \cite{SurfaceCode,TopoEntanglement}. One considers two types of loops, namely, (A) paths comprised of a closed sequence of the tensor network bonds, and (B) closed paths in the ambient manifold (outside the tensor network) that intersects only the bonds of the tensor network, such that the two bulk sites associated with the intersected bonds are located inside and outside of the loop respectively.

State $\myphisym$ is invariant under any type B loop of $\hat{Z}$'s, which follows simply from the fact that $\myphisym$ has a local $Z_2$ gauge symmetry. See \textit{Lemma 1} in Appendix \ref{app:gauge}. In addition, $\myphisym$ must be invariant under the action of type A loops of $\hat{X}$ operators
\begin{equation}\label{eq:XX}
\hat{X} \equiv \mathbb{V} \rightarrow \mathbb{V},~~~\hat{X} \equiv \ket{e}\bra{o} + \ket{o}\bra{e}.
\end{equation}
However, it can be shown that the bulk expectation value of any type A loop of $\hat{X}$'s is identically zero, see \textit{Lemma 2} in Appendix \ref{app:gauge}. This implies that the state $\myphisym$ does not have $Z_2$ topological order.

However, since the MERA representation of a quantum many-body state is generally not unique (see, for example, the discussion in Sec.~\ref{ssec:onetomany}). Is there then a different MERA representation, composed of isometric and ${Z}_2$-symmetric tensors, of the GHZ state that perhaps has $Z_2$ topological order? The answer is still no, since the proof presented in Appendix \ref{app:gauge} applies to any lifted MERA that is obtained by replacing $\ughz$ and $\wghz$ with \textit{arbitrary} isometric and ${Z}_2$-symmetric tensors.

 %%%%%%%%%%%%%%%%%%%%%%%%%%%%%%%%%%%%%%%%%%%%%%%%%%%%%%%%%%%%%%%%%%%%%%%%%%%%%%%%%%%%%%%%%%%%%%%%%%
\begin{figure}
  \includegraphics[width=\columnwidth]{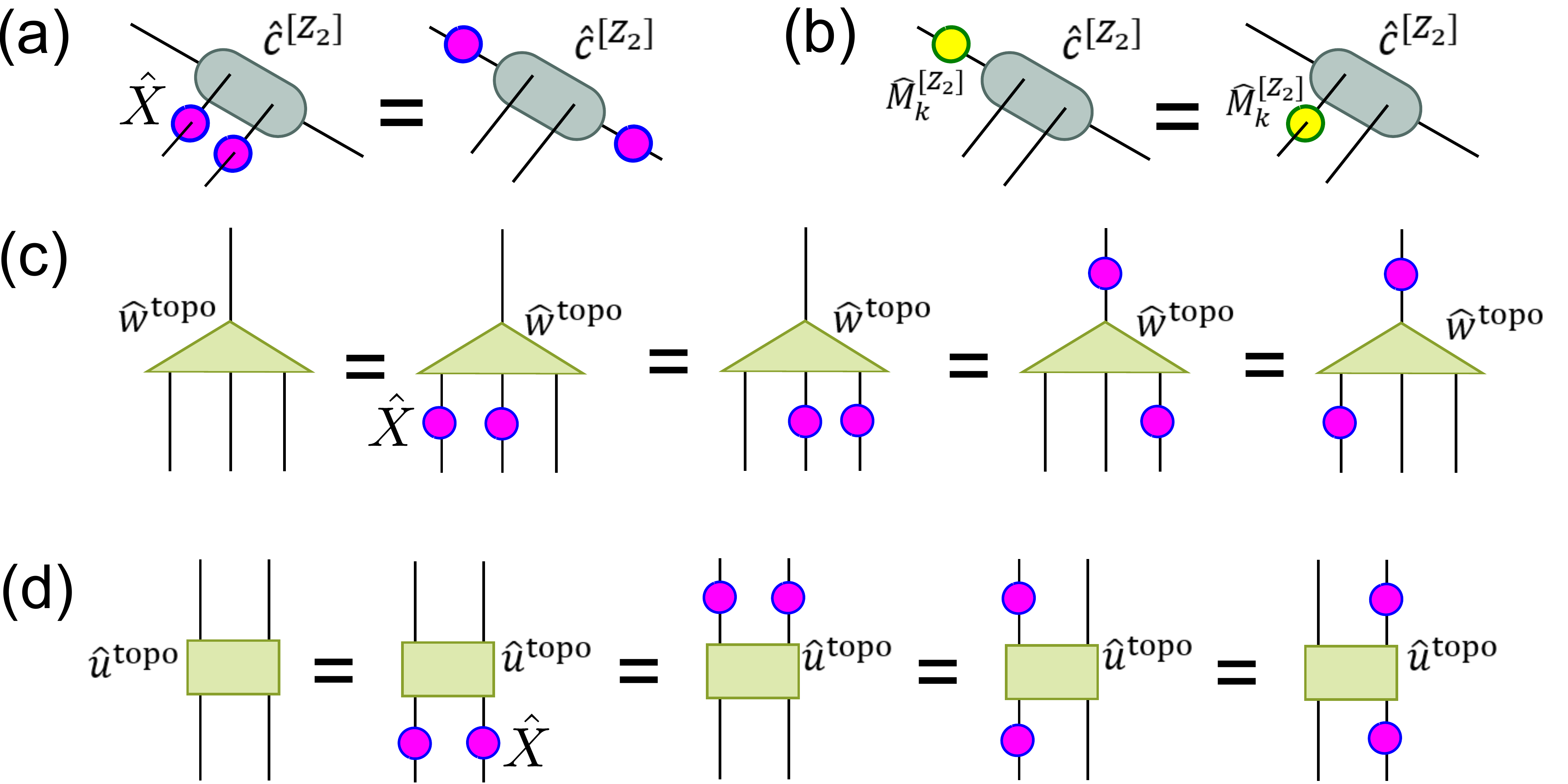}
\caption{\label{fig:topotensors} (a) $\hat{X}$ operators applied on the two open indices of the $Z_2$ copy tensor $\hat{c}^{\zz}$ [\eref{eq:z2copy}] transfers to the two bond indices. (b) A \textit{diagonal} bond transformation $\hat{M}^{\zz}_k$ applied on any of the two bond indices of the copy tensor $\hat{c}^{\zz}$ transfers to the closer open index. Here illustrated for one of the bond indices only. (c,d) Equalities illustrating that tensors $\utopo$ and $\wtopo$ remain invariant under the action of the $\hat{X}$, \eref{eq:XX}, applied on any two indices.}
\end{figure}
%%%%%%%%%%%%%%%%%%%%%%%%%%%%%%%%%%%%%%%%%%%%%%%%%%%%%%%%%%%%%%%%%%%%%%%%%%%%%%%%%%%%%%%%%%%%%%%

 %%%%%%%%%%%%%%%%%%%%%%%%%%%%%%%%%%%%%%%%%%%%%%%%%%%%%%%%%%%%%%%%%%%%%%%%%%%%%%%%%%%%%%%%%%%%%%%%%%
\begin{figure}
  \includegraphics[width=\columnwidth]{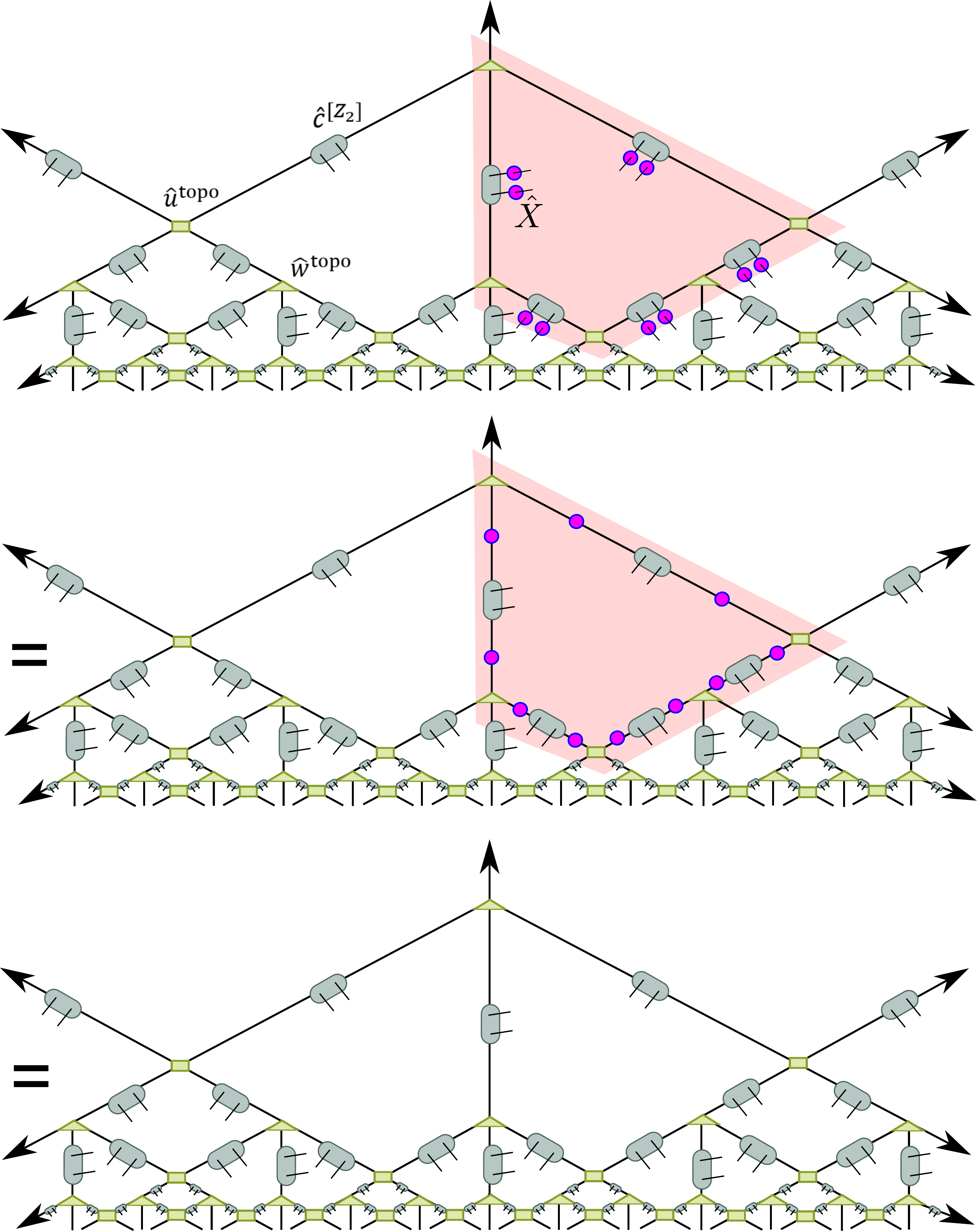}
\caption{\label{fig:gaugePlaquette} The bulk state $\myphisymtopo$ remains invariant under the action of $\hat{X}$ operators applied on bulk sites located around a plaquette on the bulk lattice, here illustrated for the action of $\hat{X}$'s on the highlighted plaquette (red). This is shown by means of two equalities. The first equality results from applying the equality depicted in \fref{fig:topotensors}(a) to all the copy tensors located around the plaquette. The second equality results from using the equalities depicted in \fref{fig:topotensors}(c)-(d), which eliminates the $\hat{X}$ operators.}
\end{figure}
%%%%%%%%%%%%%%%%%%%%%%%%%%%%%%%%%%%%%%%%%%%%%%%%%%%%%%%%%%%%%%%%%%%%%%%%%%%%%%%%%%%%%%%%%%%%%%%

Next, consider a different MERA tensor network $\mathcal{T}^{\mbox{\tiny topo}}$ comprised of copies of tensors  $\wtopo \equiv \wghz$, and $\utopo$ defined as:
 \begin{equation}\label{eq:z2MERA1}
    (\utopo)^{ij}_{kl} =
    \begin{cases}
      1, & \text{if } (i+j+k+l)\text{ mod } 2 =0 \\
      0, & \text{otherwise.}
    \end{cases}
  \end{equation}
Analogous to $\mathcal{T}$, the tensor network $\mathcal{T}^{\mbox{\tiny topo}}$ also represents the GHZ state $\mypsisym$. Namely, by contracting all the tensors of $\mathcal{T}^{\mbox{\tiny topo}}$ and changing the site basis to $\ket{\pm}$ one obtains the probability amplitudes in \eref{eq:ghz1}. On the other hand, tensor network $\mathcal{T}^{\mbox{\tiny topo}}$ cannot be obtained from $\mathcal{T}$ by applying bond transformations.

Tensor $\utopo$ is $Z_2$-symmetric since $\utopo = (\hat{Z} \otimes \hat{Z})  \utopo (\hat{Z}^\dagger \otimes \hat{Z}^\dagger)$. However, $\utopo$ is neither an isometry nor a unitary tensor. Instead, $\utopo$ is a projector,
\begin{equation}
\sum_{mn}(\utopo)^{ij}_{mn}(\utopo)^{mn}_{kl} = (\utopo)^{ij}_{kl}.
\end{equation}
[That is, $(\utopo)^2 = (\utopo)$.]

Let $\myphisymtopo$ denote the bulk state obtained by lifting $\mathcal{T}^{\mbox{\tiny topo}}$, by inserting copies of the $Z_2$-symmetric copy tensor $\hat{c}^{\zz}$ on the bonds of $\mathcal{T}^{\mbox{\tiny topo}}$. We show that $\myphisymtopo$ is the ground state of the $Z_2$ \textit{surface code} Hamiltonian, here defined on a hyperbolic lattice  \cite{SurfaceCode}. The ground state of the $Z_2$ surface code is known to have $Z_2$ topological order. Let $v$ and $p$ denote the vertices and plaquettes of the bulk hyperbolic lattice, and $s(v)$ and $s(p)$ denote the set of bulk sites immediately surrounding vertex $v$ and those located around plaquette $p$ respectively. The $Z_2$ surface code Hamiltonian is defined here as
\begin{equation}\label{eq:surfacecode}
\hat{H}^{\tiny \mbox{topo}} \equiv -\sum_{v} (\bigotimes_{i \in s(v)}\hat{Z}_i) - \sum_{p} (\bigotimes_{i \in s(p)}\hat{X}_i).
\end{equation}
The bulk state $\myphisymtopo$ is the ground state of $\hat{H}^{\tiny \mbox{topo}}$ because it remains invariant under the action of the vertex terms in \eref{eq:surfacecode} (which are simply $Z_2$ gauge transformations)---by virtue of our bulk construction---and also under the action of the plaquette terms in \eref{eq:surfacecode}, as illustrated in \fref{fig:gaugePlaquette}.

Since each pair of bulk sites associated with a bond are effectively supported only on a qubit subspace, by virtue of the gauge symmetry, our bulk state is indeed stabilised by a set of vertex and plaquette operators equivalent to the surface code state. As the state $\myphisym$ has no further symmetries or additional degrees of freedom then this is the same phase as the surface code: a phase with $Z_2$ topological order. We also refer the reader to Ref.~\onlinecite{exactTopo} where the topological properties of the lifted tensor network, which represents state $\myphisymtopo$, are explicitly demonstrated. (In Ref.~\onlinecite{exactTopo} the authors consider a 3-index copy tensor, which is isomorphic to the $Z_2$-symmetric copy tensor $\hat{c}^{\zz}$ after each pair of bond sites is projected to an effective qubit space.)

The bulk state represented by the lifted $\mathcal{T}^{\mbox{\tiny topo}}$ tensor network turned out to have topological order because the tensor network contains \textit{non-isometric} tensors, and thus avoids the argument for the absence of bulk topological order presented in Appendix \ref{app:gauge}. Thus, this example reveals an interesting interplay between the presence of an Abelian topological order in the bulk and the isometric constraints that are usually imposed on the MERA tensors. On the other hand, for non-Abelian symmetries, isometric tensors may be compatible with the presence of a bulk topological order. 
%We also remark that bulk states obtained by lifting a MERA that contains non-isometric tensors no longer necessarily exhibit the bulk features listed $(i)$-$(iii)$ in Sec.~\ref{sec:copybulkstate}.
 
%00000000000000
\subsection{A possible correspondence between 1D symmetry breaking phases and 2D topological phases?}
Viewed as ground states of local Hamiltonians, the quantum many-body states $\mypsisym$ and $\myphisymtopo$, described above, belong to two different quantum phases of matter: the boundary state $\mypsisym$ (a GHZ state) belongs to a quantum phase with spontaneously broken $Z_2$ symmetry, while the bulk state $\myphisymtopo$ belongs to a quantum phase with $Z_2$ topological order.
(In a $Z_2$ symmetry broken phase, the ground state is 2-fold degenerate. The only $Z_2$-symmetric ground states are GHZ type states dressed with local entanglement, see e.g. Appendix C in Ref.~\onlinecite{GHZSingh}.)

The $Z_2$-symmetric ground states at the RG fixed point in the phase are exactly GHZ states. Recall that the MERA representation of a ground state describes the RG flow of the ground state to a fixed point wavefunction, which is characteristic of the quantum phase to which the ground state belongs \cite{Singh131}. If the ground state belongs to a $Z_2$ symmetry broken phase, and if we target the $Z_2$-symmetric ground subspace by protecting the symmetry along the RG flow (employing $Z_2$-symmetric tensors) then the ground state flows to a GHZ state.
%(This is because the only $Z_2$-symmetric ground states in a $Z_2$ symmetry broken phase are GHZ states dressed with local entanglement, see e.g. Appendix C in Ref.~\onlinecite{GHZSingh}.)

Subsequently, if the fixed point tensors in the MERA representation of a $Z_2$-symmetry broken ground state are, in fact, $\utopo$ and $\wtopo$ (which represent a GHZ state) it is tempting to conclude that our bulk construction leads to an \textit{emergent} $Z_2$ topological order at the RG fixed point in a 1D $Z_2$ symmetry broken phase. Here by emergent we mean that we can systematically obtain a quantum state with $Z_2$ topological order from the MERA representationof a $Z_2$ symmetry broken ground state (by lifting the MERA).
However, this argument does not quite work since the MERA representation of the GHZ state is not unique. For example, as mentioned previously, the GHZ state may also be represented by the MERA $\mathcal{T}$, which does not lift to a topologically ordered bulk state. 

On the other hand, we have isolated a condition under which our holographic correspondence could lead to an emergent \textit{Abelian} topological order in the bulk. Namely, if non-isometric tensors are permitted, and in fact preferred, in the MERA representation of a ground state, while protecting the symmetry along the RG flow. In this case, the fixed point tensors in a $Z_2$-broken phase are indeed given by \eref{eq:z2MERA}, and therefore one can argue for an emergent topological order as described above.

 %%%%%%%%%%%%%%%%%%%%%%%%%%%%%%%%%%%%%%%%%%%%%%%%%%%%%%%%%%%%%%%%%%%%%%%%%%%%%%%%%%%%%%%%%%%%%%%%%
\begin{figure}[t]
  \includegraphics[width=\columnwidth]{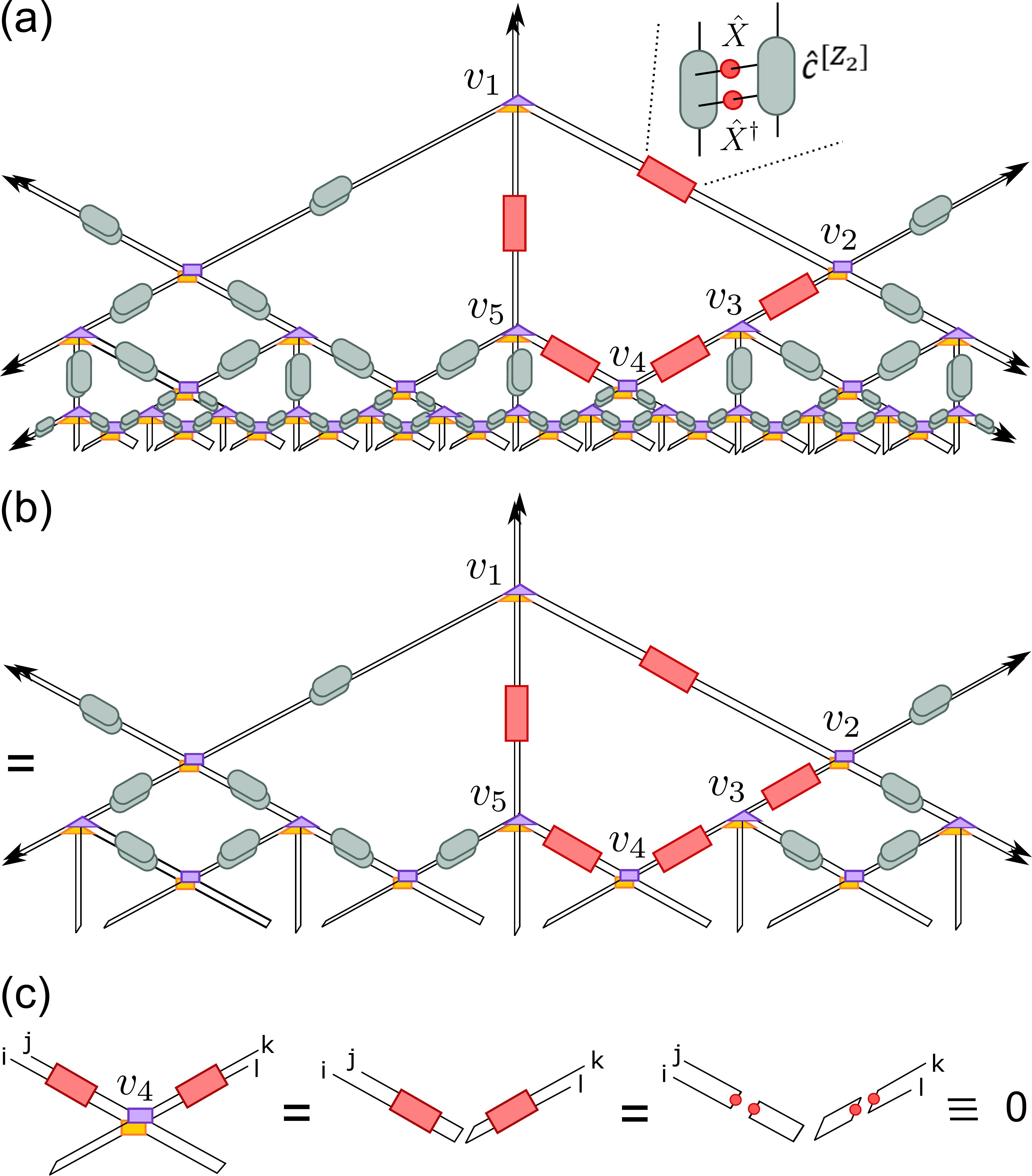}
\caption{\label{fig:bulkLoop} (a) Tensor network contraction equating to the bulk expectation value of a loop of $\hat{X}$'s. (b) The tensor network contraction resulting from simplifying the tensors located below the loop. First, the tensors located at the bottom are contracted with their adjoints and thus cancel out. Consequently, pairs of copy tensors are seen to be contracted together and also cancel out, thanks to the equality depicted \fref{fig:copyTensor}$(b)$. (c) Zoom in to the contraction around tensor $v_4$. Tensor $\hat{u}$ cancels with its adjoint, resulting in two separate contractions. Each of these is identically zero since the trace of $\hat{X}$ is 0.}
\end{figure}
%%%%%%%%%%%%%%%%%%%%%%%%%%%%%%%%%%%%%%%%%%%%%%%%%%%%%%%%%%%%%%%%%%%%%%%%%%%%%%%%%%%%%%%%%%%%%%%
%%%%%%%%%%%%%%%%%%%%%%%%%%%%%%%%%%%%%%%%

%%%%%%%%%%%%%%%%%%%%%%%%%%%%%%%%%%%%%%%%%%%%%%%%%%%%%%%%%%%%%%
\subsection{Proofs}\label{app:gauge}

In this section, we prove two lemmas that were used in the discussion presented in this Appendix.

Consider a vector space $\mathbb{V} \cong \mathbb{C}_2$ that is equipped with the action of the group $Z_2=\{\hat{I},\hat{Z}\}$. The group acts on the space $\mathbb{V}$ by means of the unitary representation
$\hat{I} = \smallmat{1}{0}{0}{1},~\hat{Z} = \smallmat{1}{0}{0}{-1}$. Under the action of the symmetry, the space $\mathbb{V}$ decomposes as $\mathbb{V} \cong \mathbb{V}_e \oplus \mathbb{V}_o$
where $\mathbb{V}_e$ and $\mathbb{V}_o$ are the two irreps of $Z_2$. Denote by $\ket{e}$ and $\ket{o}$ a basis in the one dimensional vector spaces $\mathbb{V}_e$ and $\mathbb{V}_o$. Also, define the $Z_2$ irrep flip operator $\hat{X} \equiv \ket{e}\bra{o} +\ket{o}\bra{e}$. Consider a MERA tensor network $\mathcal{T}$ composed from \textit{arbitrary} isometric and $Z_2$-symmetric tensors $\hat{u}:\mathbb{V} \otimes \mathbb{V} \rightarrow \mathbb{V} \otimes \mathbb{V}$ and $\hat{w}: \mathbb{V} \rightarrow \mathbb{V} \otimes \mathbb{V} \otimes \mathbb{V}$. Let $\mathcal{T}'$ denote the lifted MERA obtained by inserting the $Z_2$-symmetric copy tensor, \eref{eq:copycomponents}, on the bonds of $\mathcal{T}$.

\textit{Lemma 1.} Consider a loop $\mathcal{C}$ in the ambient manifold, in which the (lifted) MERA is embedded, (i) that intersects only copy tensors, and (ii) the two bulk sites associated with the open indices of an intersected copy tensor are located inside and outside of the loop respectively. Also, consider the loop operator $\hat{Z}_{\mathcal{C}} \equiv \bigotimes_{i} \hat{Z}_i$ that acts on all bulk sites $i$ located immediately inside loop $\mathcal{C}$. The lifted MERA $\mathcal{T}'$, and thus the bulk state it represents, is invariant under the action of $\hat{Z}_{\mathcal{C}}$.

\textit{Proof.} As an illustration of the general proof, consider the specific loop operator applied along the loop enclosing the tensors $v_1,v_2,v_3,v_4$ and $v_5$, depicted in \fref{fig:zLoop}. Let us apply $Z_2$ gauge transformations simultaneously around all the 5 tensors. The gauge transformations leave the lifted tensor network $\mathcal{T}'$ invariant, of course. However, the action of this gauge transformation is equivalent to applying a loop of $\hat{Z}$'s. This follows from using the equalities depicted in \fref{fig:zLoop}(a); all $\hat{Z}$ operators except those located along the loop are eliminated. Thus, this loop operator leaves $\mathcal{T}'$ invariant. This proof is readily generalized for an arbitrary loop $\mathcal{C}$, homologous to the loop considered above. $\square$

\textit{Lemma 2.} Consider a loop $\tilde{\mathcal{C}}$ comprised of a closed sequence of the MERA bonds. Also, consider the loop operator $\hat{X}_{\tilde{\mathcal{C}}} \equiv \bigotimes_{i} \hat{X}_i$ that acts on all bulk sites $i$ located along the loop $\tilde{\mathcal{C}}$. The expectation value of $\hat{X}_{\tilde{\mathcal{C}}}$ obtained from $\mathcal{T}'$ is identically zero. 

\textit{Proof.} Once again we only give an illustration of the general proof here. Consider the expectation value of a loop of $\hat{X}$'s, obtained from the lifted MERA $\mathcal{T}'$, around the tensors $v_1,v_2,v_3,v_4$ and $v_5$ depicted in \fref{fig:bulkLoop}{(a)}. The tensor network contraction equating to the expectation value is illustrated in the figure. The contraction can be simplified by iteratively applying two sequences of cancellations, proceeding upwards from the boundary. First, the tensors located at the very bottom are contracted with their adjoints and thus cancel out. Consequently, pairs of copy tensors are seen to be contracted together and also cancel out, thanks to the equality depicted \fref{fig:copyTensor}{(b)}. By applying these two simplifications iteratively most tensors below the loop cancel out, and we are left with the contraction depicted in \fref{fig:bulkLoop}{(b)}.

(The contraction depicted in \fref{fig:bulkLoop}{(a)} is a simple illustration where only one layer of tensors appears below the loop. More generally, the loop may appear deep in the bulk, located above many layers of tensors. But by iteratively applying the simplifications described above such a contraction reduces, once again, to the contraction depicted in \fref{fig:bulkLoop}{(b)}.)

Next, consider the tensor contractions around tensor $v_4$, which is separately depicted in \fref{fig:bulkLoop}{(c)}. Since tensor $\hat{u}$ is isometric, it cancels out when contracted with its adjoint, leading to the first equality depicted on the left in \fref{fig:bulkLoop}{(c)}. Each of the resulting two contractions (shown in the middle in \fref{fig:bulkLoop}{(c)}) consists of two bond tensors and an $\hat{X}$ operator, and is identically zero since the trace of $\hat{X}$ is 0. Thus, the expectation value of the loop operator is identically zero. $\square$

%%%%%%%%%%%%%%%%%%%%%%%%%%%%%%%%%%%%%%%%%%%%%%%%%%%%%%%%%%%%%%%%%%%%%%%%%%%%%%%%%%%%%%%%%%%%%%%
%%%%%%%%%%%%%%%%%%%%%%%%%%%%%%%%%%%%%%%%
\section{Schmidt decomposition of a bulk state}\label{app:areaLaw}
The \textit{Schmidt decomposition} of a quantum state belonging to a bipartite tensor product space $\mathbb{V}^{(A)} \otimes \mathbb{V}^{(B)}$ is
\begin{equation}\label{eq:schmidt}
\ket{\Psi} = \sum_{\alpha = 1}^{n} \mu_{\alpha} \ket{\Omega^{[A]}_\alpha} \otimes \ket{\Omega^{[B]}_\alpha},
\end{equation}
where $\mu_{\alpha} > 0$ are the \textit{Schmidt coefficients}, and $\{\ket{\Omega^{[A]}}_\alpha\}$ and $\{\ket{\Omega^{[B]}}_\alpha\}$ is an orthonormal basis in spaces $\mathbb{V}^{(A)}$ and $\mathbb{V}^{(B)}$ respectively. The decomposition \ref{eq:schmidt} is useful since the reduced density matrix of  the parts $A$ and $B$ is diagonal in the \textit{Schmidt basis}, namely,
\begin{equation}\label{eq:schmidtrho}
\begin{split}
\hat{\rho}^{[A]} &\equiv \sum_{\alpha = 1}^{n} \mu^2_{\alpha}~\ket{\Omega^{[A]}_\alpha}\bra{\Omega^{[A]}_\alpha},\\
\hat{\rho}^{[B]} &\equiv \sum_{\alpha = 1}^{n} \mu^2_{\alpha} ~\ket{\Omega^{[B]}_\alpha}\bra{\Omega^{[B]}_\alpha}.
\end{split}
\end{equation}
In particular, the rank of the reduced density matrices $\hat{\rho}^{[A]}$ and $\hat{\rho}^{[B]}$ is equal to $n$.

In this appendix, we derive a Schmidt decomposition for a bulk state $\myphi$. We will use this Schmidt decomposition to derive two results: $(i)$ the bulk state exhibits an area law scaling of entanglement, and $(ii)$ $\myphi$ can be viewed as the ground state of a local, gauge-invariant Hamiltonian.

%%%%%%%%%%%%%%%%%%%%%%%%%%%%%%%%%%%%%%%%%%%%%%%%%%%%%%%%%%%%%%%%%%%%%%%%%%%%%%%%%%%%%%%%%%%%%%%%%%
\begin{figure}
  \includegraphics[width=\columnwidth]{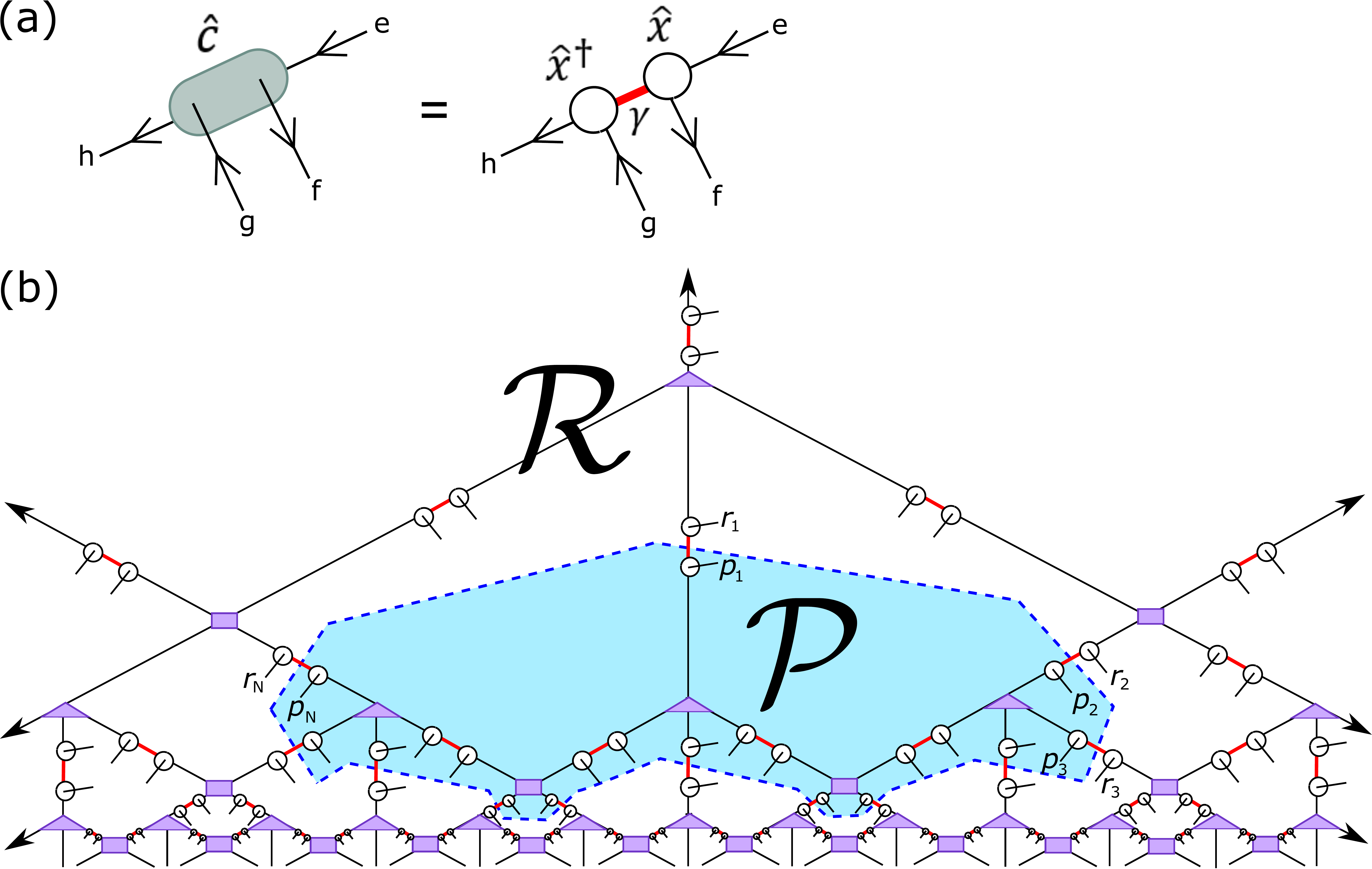}
\caption{\label{fig:schmidtFull} (a) The graphical representation of the decomposition the $\mathcal{G}$-symmetric copy tensor in terms of a trivalent tensor $\hat{x}$, \eref{eq:copydecompose}, exposing an intermediate bond index $\gamma$ (red) that carries the trivial symmetry charge. (b) A bipartition of the lifted MERA into parts $\mathcal{P}$ and $\mathcal{R}$ by a path (dashed contour) that only intersects (the red bonds of) $N$ copy tensors. $\{p_1,p_2,\ldots, p_N\}$ and $\{r_1,r_2,\ldots, r_N\}$ denote the open indices of the $N$ intersected copy tensors.}
\end{figure}
%%%%%%%%%%%%%%%%%%%%%%%%%%%%%%%%%%%%%%%%%%%%%%%%%%%%%%%%%%%%%%%%%%%%%%%%%%%%%%%%%%%%%%%%%%%%%%%%%%

For our purposes, we decompose the copy tensor $\hat{c}$, \eref{eq:copycomponents}, in terms of a 3-index, $\mathcal{G}$-symmetric tensor $\hat{x}$
\begin{equation}\label{eq:copydecompose}
(\hat{c})^{ef}_{gh} = \sum_{\gamma} (\hat{x})^{e}_{\gamma g} (\hat{x}^\dagger)^{\gamma f}_{h},
\end{equation}
depicted in \fref{fig:schmidtFull}. In the irrep basis,
\begin{equation}\label{eq:symBasis}
\begin{split}
\ket{e} &\equiv \ket{a,t_a,m_a},~~~~\ket{f} \equiv \ket{b,t_b,m_b},\\ 
\ket{g} &\equiv \ket{c,t_c,m_c},~~~~~\ket{h} \equiv \ket{d,t_d,m_d},
\end{split}
\end{equation}
the index values that correpond to non-zero components of $\hat{c}$ satisfy
\begin{equation}
\begin{split}
&a = b = c = d,\\
&m_a = m_b,~~m_c = m_d,\\
&t_a = t_b = t_c = t_d.
\end{split}
\end{equation}
The intermediate index $\gamma \equiv (0, t_{\gamma})$ carries only the trivial charge and takes $\sum_a d_a$ number of values. We establish a one-to-one correspondence between the index $\gamma$ and the degeneracy index $(a,t_a)$ and denote it as
 $\gamma \leftrightarrow \gamma(a,t_a)$.

In the irrep basis, tensor $\hat{x}$ decomposes as (Wigner-Eckart theorem)
\begin{equation}
\hat{x} \cong \bigoplus_a (\hat{x}^{\mbox{\tiny deg}}_a \otimes \hat{I}_{\eta_a}),
\end{equation}
where the only non-zero components of the degeneracy tensors $\hat{x}^{\mbox{\tiny deg}}_a$ are
\begin{equation}
(\hat{x}^{\mbox{\tiny deg}}_a)^{t_a}_{\gamma(a,t_a),t_a} = \frac{1}{\proot{4}{\eta_a}},~~~\mbox{for all}~t_a \in \{1,2,\ldots,d_a\}.
\end{equation}

Next, consider a closed path on the ambient manifold---into which the (lifted) MERA is embedded---that intersects only the new bonds resulting from decomposing the copy tensors, as illustrated by the dashed contour depicted in \fref{fig:schmidtFull}. Such a path bipartitions the lifted MERA, and the bulk lattice $\mathcal{M}$, into parts $\mathcal{P}$ and $\mathcal{R}$. The two bulk sites that are associated with each intersected copy tensor are split amongst parts $\mathcal{P}$ and $\mathcal{R}$ respectively.

Let $\hat{P}$ and $\hat{R}$ denote the tensors obtained by contracting all the tensors located in parts $\mathcal{P}$ and $\mathcal{R}$ respectively. Also, let $p \equiv (p_1,p_2,\ldots,p_N,p_{N+1},\ldots)$ and $r \equiv (r_1,r_2,\ldots,r_N,r_{N+1},\ldots)$ denote the tuple of all open indices located in $\mathcal{P}$ and $\mathcal{R}$ respectively.

The first $N$ indices--- $\{p_1,p_2,\ldots, p_N\}$ and $\{r_1,r_2,\ldots, r_N\}$---are the open indices of the $N$ intersected copy tensors, as illustrated in \fref{fig:schmidtFull}. Analogously, let $\alpha \equiv (\gamma_1,\gamma_2,\ldots,\gamma_N)$ denote the tuple of all (red) indices that connect part $\mathcal{P}$ with $\mathcal{R}$, see \fref{fig:schmidtFull}.

The bulk state $\myphi$ can be expressed as 
\begin{equation} \label{eq:schmidtbulk}
\myphi = \sum_{\alpha} \ket{\tilde{\Omega}^{[\mathcal{P}]}_{\alpha}} \otimes \ket{\tilde{\Omega}^{[\mathcal{R}]}_{\alpha}},
\end{equation}
where the vectors $\{\ket{\tilde{\Omega}^{[\mathcal{P}]}_{\alpha}}\}$ and $\{\ket{\tilde{\Omega}^{[\mathcal{R}]}_{\alpha}}\}$ are given by
\begin{equation}\label{eq:basisPQ}
\ket{\tilde{\Omega}^{[\mathcal{P}]}_{\alpha}} \equiv \sum_{p} \hat{P}_{p\alpha}  \ket{p},~~~~~\ket{\tilde{\Omega}^{[\mathcal{R}]}_{\alpha}} \equiv \sum_{r} \hat{R}_{\alpha r} \ket{r},
\end{equation}
and $\ket{p} \equiv  \ket{p_1} \otimes \ket{p_2} \otimes \ldots$ and $\ket{r} \equiv  \ket{r_1} \otimes \ket{r_2} \otimes \ldots$. These vectors form an orthogonal basis for the subsystems $\mathcal{P}$ and $\mathcal{R}$ respectively, namely,
\begin{equation}\label{eq:orthorgonalbasis}
\braket{\tilde{\Omega}^{[\mathcal{P}]}_{\alpha}}{\tilde{\Omega}^{[\mathcal{P}]}_{\alpha'}} \propto \delta_{\alpha \alpha'}~~~\mbox{and}~~~\braket{\tilde{\Omega}^{[\mathcal{R}]}_{\alpha}}{\tilde{\Omega}^{[\mathcal{R}]}_{\alpha'}} \propto \delta_{\alpha \alpha'}.
\end{equation}
This can be understood as follows. The one-to-one map $\gamma(a,t_a)$ also corresponds to a one-to-one map between the kets $\ket{\gamma} \leftrightarrow \ket{a,t_a}$. Two different tuples $\alpha$ and $\alpha'$ differ in some entry $\gamma_i$, and therefore also correspond to different elements of the tensor product basis $\ket{p}$ and $\ket{p'}$ (and also $\ket{r}$ and $\ket{r'}$). However, $\bra{p}p'\rangle = \delta_{p,p'}$, thus leading to \eref{eq:orthorgonalbasis}.

Let $\eta^{(\mathcal{P})}_{\alpha}$ and $\eta^{(\mathcal{R})}_{\alpha}$ denote the norm of the vectors $\ket{\tilde{\Omega}^{[\mathcal{P}]}_{\alpha}}$ and $\ket{\tilde{\Omega}^{[\mathcal{R}]}_{\alpha}}$ respectively. We write
\begin{equation}
\ket{\tilde{\Omega}^{[\mathcal{P}]}_{\alpha}} = \eta^{(\mathcal{P})}_{\alpha}\ket{{\Omega}^{[\mathcal{P}]}_{\alpha}},~~~\ket{\tilde{\Omega}^{[\mathcal{R}]}_{\alpha}} = \eta^{(\mathcal{R})}_{\alpha}\ket{{\Omega}^{[\mathcal{R}]}_{\alpha}},
\end{equation}
where $\{\ket{{\Omega}^{[\mathcal{P}]}_{\alpha}}\}$ and $\{\ket{{\Omega}^{[\mathcal{R}]}_{\alpha}}\}$ denote the normalized \textit{Schmidt basis} in parts $\mathcal{P}$ and $\mathcal{R}$ respectively.
Reorganizing \eref{eq:schmidtbulk} we obtain the \textit{Schmidt decomposition} of the bulk state for the bipartition $\mathcal{P}:\mathcal{R}$,
\begin{equation} \label{eq:schmidtbulk}
\myphi = \sum_{\alpha} (\eta^{[\mathcal{P}]}_{\alpha} \eta^{[\mathcal{R}]}_{\alpha})~\ket{{\Omega}^{[\mathcal{P}]}_{\alpha}} \otimes \ket{{\Omega}^{[\mathcal{R}]}_{\alpha}},
\end{equation}
where $\eta^{[\mathcal{P}]}_{\alpha} \eta^{[\mathcal{R}]}_{\alpha} > 0$ are the \textit{Schmidt coefficients} that appear in \eref{eq:schmidt}. It is notable that the Schmidt basis in a region, say $\mathcal{R}$, is obtained by simply contracting all the tensors in the region, \eref{eq:basisPQ}.

 %%%%%%%%%%%%%%%%%%%%%%%%%%%%%%%%%%%%%%%%%%%%%%%%%%%%%%%%%%%%%%%%%%%%%%%%%%%%%%%%%%%%%%%%%%%%%%%%%
\begin{figure}[t]
  \includegraphics[width=\columnwidth]{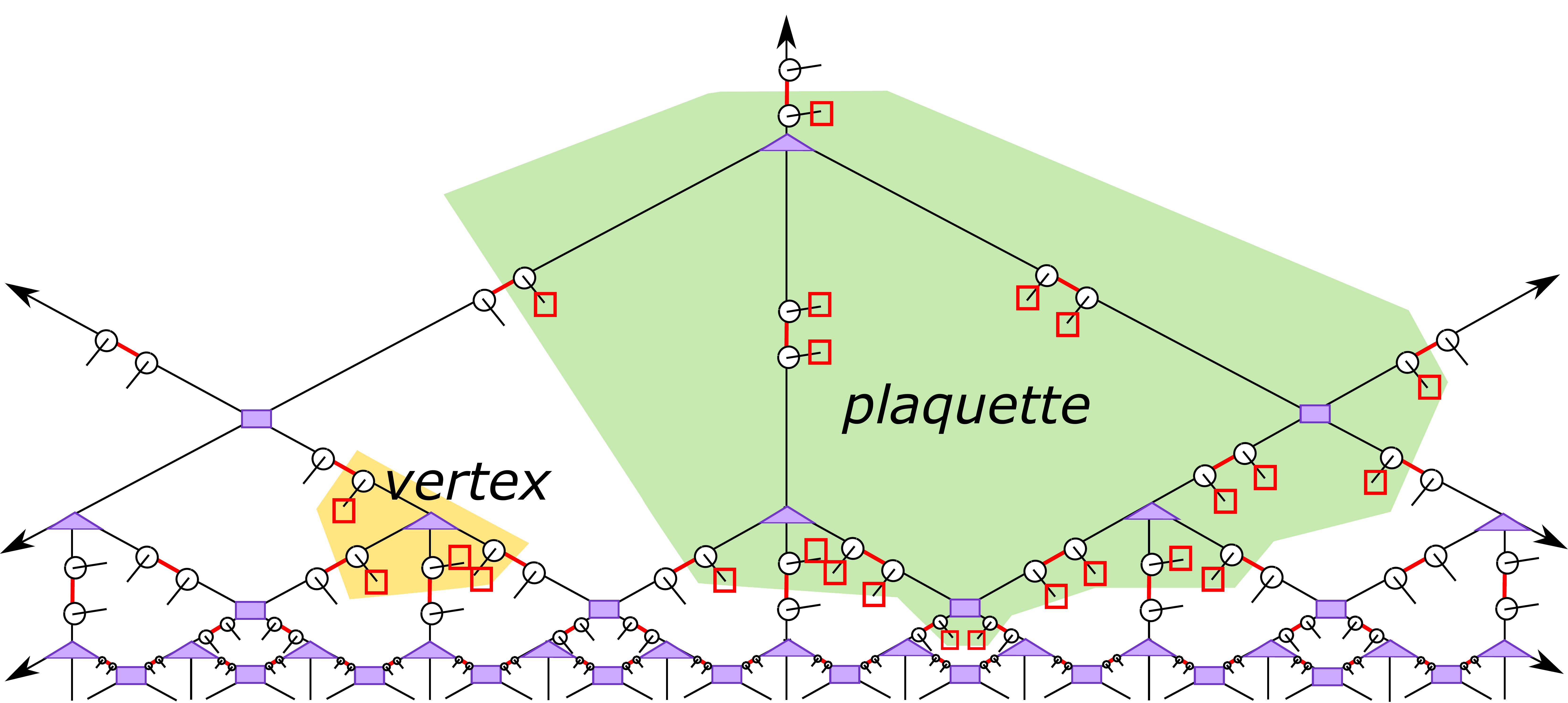}
\caption{\label{fig:parent} The bulk sites (red squares) acted on by the vertex operator $\hat{h}^{[v]}$ and the plaquette operator $\hat{h}^{[p]}$ , which appear in \eref{eq:parentHam}, are highlighted yellow and green respectively.}
\end{figure}
%%%%%%%%%%%%%%%%%%%%%%%%%%%%%%%%%%%%%%%%%%%%%%%%%%%%%%%%%%%%%%%%%%%%%%%%%%%%%%%%%%%%%%%%%%%%%%%

Equation (\ref{eq:schmidtbulk}) implies that the rank of the reduced density matrix $\hat{\rho}^{[\mathcal{P}]}$ is at most equal to the number of different values the tuple $\alpha$ assumes, which equals $(\sum_a d_a)^N$. This implies that the entanglement entropy $S(\hat{\rho}^{[\mathcal{P}]}) = -\mbox{Tr} (\hat{\rho}^{[\mathcal{P}]} \mbox{log}~\hat{\rho}^{[\mathcal{P}]})$ of subsystem $\mathcal{P}$ is proportional to $N$, the number of sites at the boundary of $\mathcal{P}$. Thus, the entanglement of a bulk subsystem $\mathcal{P}$ scales as the perimeter of the subsystem, often called `area law entanglement scaling' in condensed matter physics \cite{AreaLaw}. Here, we have bipartitioned the tensor network in a particular way, which may give the impression that the area law entanglement proved above is exhibited only by such regions. However, the above argument is only an illustration of the more general result proved in Ref.~\onlinecite{TNC}, namely, area law entanglement is exhibited by any bulk region.

%%%%%%%%%%%%%%%%%%%%%%%%%%%%%%%%%%%%%%%%%%%
\section{A gauge-invariant parent Hamiltonian for a bulk state}\label{app:parentHam}
Consider the Hamiltonian
\begin{equation}\label{eq:parentHam}
\hat{H}^{\tiny \mbox{bulk}} = -\sum_{v} \hat{h}^{[v]} - \sum_{p} \hat{h}^{[p]},
\end{equation}
where $v$ and $p$ labels the tensors and plaquettes of the MERA tensor network. Here $\hat{h}^{[v]}$ is the projector on to the support of the reduced density matrix of bulk sites located immediately around vertex $v$ (highlighted yellow in \fref{fig:parent}). Analogously, $\hat{h}^{[p]}$ is the projector on to the support of the reduced density matrix of bulk sites located along and immediately surrounding plaquette $p$ (highlighted green in \fref{fig:parent}) respectively. We have
\begin{equation}
\hat{h}^{[v]} \equiv \sum_{\alpha} \ket{{\Omega}^{[v]}_{\alpha}}\bra{{\Omega}^{[v]}_{\alpha}},~~~\hat{h}^{[p]} \equiv \sum_{\alpha} \ket{{\Omega}^{[p]}_{\alpha}}\bra{{\Omega}^{[p]}_{\alpha}},
\end{equation}
where $\{\ket{{\Omega}^{[v]}_{\alpha}}\}$ and $\{\ket{{\Omega}^{[p]}_{\alpha}}\}$ are the Schmidt bases for the vertex and plaquette sites respectively. These bases can be obtained from the tensors of the lifted MERA in a simple way, as described in Appendix \ref{app:areaLaw}. State $\myphi$ is a ground state of the Hamiltonian $\hat{H}^{\tiny \mbox{bulk}}$ since
\begin{equation} \label{eq:parentHam1}
\begin{split}
\myphic \hat{h}^{[v]} \myphi &= \mbox{tr}(\hat{\rho}^{[v]} \hat{h}^{[v]}) = 1,\\
\myphic \hat{h}^{[p]} \myphi &=  \mbox{tr}(\hat{\rho}^{[p]} \hat{h}^{[p]}) = 1~~~\forall v,p.
\end{split}
\end{equation}
The Hamiltonian $\hat{H}^{\tiny \mbox{bulk}}$ is \textit{local} since the vertex and plaquette terms act on 4 bulk sites and 20 bulk sites respectively.  It is readily checked that the Hamiltonian $\hat{H}^{\tiny \mbox{bulk}}$ is also gauge-invariant, since both $\hat{h}^{[v]}$ and $\hat{h}^{[p]}$ commute with the gauge transformations.

\section{Examples of gauge-invariant operators for spin network states}\label{app:holonomy}
In this appendix, we recall the definition of gauge field holonomy operators, which are a natural choice for gauge-invariant observables in the bulk. 
For a discrete group $\myg$, the \textit{group-valued} holonomy around the counterclockwise oriented closed path $\partial f$ bounding a contiguous region $f$ on the lattice is
\begin{equation}
h_f=\prod_{e\in\partial f}g^{o(e,f)}_e,
\end{equation}
where $g_e\in \myg$ is the group element on edge $e$, the ordered product is taken along the boundary $\partial f$ and $o(e,f)= 1$ if the edge $e$ is oriented the same direction as $f$ and $o(e,f)= -1$ if the edge $e$ is oriented the opposite direction as $f$. For continuous lie groups and in curved spacetime, the group element on an edge is obtained as is done in lattice gauge theory. Take the path ordered integral of the gauge field $\hat{A}(x)$ valued in $\mathfrak{g}$ over the edge $e$:
\begin{equation}
h_{e}=\mathcal{P}e^{-i\Lambda\oint_{e} g_{\mu\nu}(x)\hat{A}^{\mu}(x)dx^{\nu}},
\end{equation}
where $g_{\mu\nu}(x)$ is the metric, and $\Lambda$ is the gauge field coupling strength \cite{CurvedHolonomy}. 

In our case, we have a description of bulk state in terms of a spin network basis in the gauge-invariant sector of the bulk Hilbert space. An edge of the bulk lattice with irrep label $j$ is equivalent to the reversed oriented edge labelled by the conjugate irrep $j^*$. In the spin network basis the holonomy operator is given by
\begin{equation}
\hat{W}(h_p)=\sum_j\tr[D^{(j)}(h_p)]\hat{B}^j_p,
\end{equation}
where $D^{(j)}(h)$ is the irrep $j$ of the group element $h \in \myg$, and the sum is taken over all irrep labels. The operator $\hat{B}^j_p$ inserts a flux of type $j$ into the plaquette and its matrix elements in the spin network basis can be determined using recoupling formulae, see e.g. Ref.~\onlinecite{StringNet}.

Generically, consider a plaquette which is an $n$-gon with boundary edges carrying irreps $j_1,j_2,\ldots j_n$, and incident edges to the vertices labelled by $k_1,k_2,\ldots k_n$. The spin network for this state can be labelled $\ket{\Phi;k_1,\ldots k_n;j_1,\ldots j_n}$ where $\Phi$ indicates the configurations of all the other edges not touching the plaquette. Assuming for simplicity that all boundary edges have the same orientation as $p$ and all incident edges are directed toward the vertices of $p$, the matrix elements for the plaquette flux operator are:

\begin{equation}
\begin{split}
&\bra{\Phi;k'_1,\ldots k'_n;j'_1,\ldots j'_n}\hat{B}^s_p \ket{\Phi;k_1,\ldots k_n;j_1,\ldots j_n}=\\
& \left[\prod_{l=1,r=1}^n \delta_{k_r,k'_r}\sqrt{\frac{d_{j_{\ell}^{'}}}{d_{k_{\ell}}}} \right] \left(F^{j_n^*, j_1, j_1^{'*}}_{j_{n}^{'}}\right)^{*}_{k^*_{1},s^*} \left(F^{j_1^*, j_2, j_2^{'*}}_{j_{1}^{'}}\right)^{*}_{k^*_2,s^*}
\\
& \ldots
\left(F^{j_{n-2}^*, j_{n-1}, j_{n-1}^{'*}}_{j_{n-2}^{'}}\right)^{*}_{k^*_{n-1},s^*} \left(F^{j_{n-1}^*, j_n, j_n^{'*}}_{j_{n-1}^{'}}\right)^{*}_{k^*_n,s^*}
\end{split}
\end{equation} 
where the $F$'s are recoupling coefficients (6-j symbols) that describe coupling the three irreps (labelled by superscripts) to a fourth total irrep (labelled by subscript).
For other orientations of the edges simply replace the labels of the reverse oriented edges in the expressions above by their conjugates.

Let us consider specific examples.

\subsection{Abelian gauge groups}
\emph{The cyclic group $\mathbb{Z}_d$.---} For gauge group $\myg=\mathbb{Z}_d$ the irreps $j\in\{0,1,\ldots d-1\}$ are in one to one correspondence with the group elements taking values in $\mathbb{Z}_d$, and the conjugate irrep satisfies $j^*=-j=d-j$. The gauge field configurations are possibly overlapping closed loops of flux, configurations termed string nets. The plaquette operator is
\begin{equation}
\hat{B}^j_p=\prod_{e\in \partial p}X^{o(e,p)j}_e
\end{equation}
where $o(e,p)= 1 (-1)$ if the edge $e$ is oriented the same direction (opposite direction) to the orientation of the plaquette boundary $\partial p$. Plaquettes can be give a uniform orientation which we choose to be counterclockwise. Here $\hat{X}^j=\sum_{k=0}^{d-1}\ket{k\oplus_d j}\bra{k}$, where $\oplus_d$ is addition modulo $d$.
The group $\mathbb{Z}_2$ is particularly simple having self conjugate irreps meaning the only allowed gauge field configurations are non intersecting loops yielding $\hat{B}^0_p ={\bf 1}$ and $\hat{B}^0_p =\prod_{e\in \partial p}\hat{X}$.

\emph{The group $U(1)$.---} For the gauge group $\myg = U(1)$, the irreps are labelled by integers, $j \in \mathbb{Z}$, and the conjugate irrep satisfies $j^*=-j$. Consider a vertex of the spin network with two incoming edges labeled by irreps $j_1$ and $j_2$ and an outgoing edge labelled $j_3$. These must satisfy the branching rule $j_1+j_2-j_3=0$ implying only closed strings of irreps or sums thereof appear on the network. The plaquette operator is
\begin{equation}
\hat{B}^j_p=\prod_{e\in \partial p}\hat{L}^{o(e,p)j}_e
\end{equation}
where $\hat{L}^j=\sum_{k\in \mathbb{Z}}\ket{k+j}\bra{k}$. These operators are infinite dimensional but when the irreps appearing in the spin network are truncated so that $j_{\rm min}\leq j\leq j_{\rm max}$, then one can use finite dimensional truncations of $\hat{L}^j$.

%Note that, in our construction, the irreps that appear on the bonds of the holographic spin networks are already truncated, since they are carried over from the MERA representation of a $\mathcal{G}$-symmetric ground state. (The bond irreps are systematically truncated in practical MERA simulations \cite{SinghSU2,SinghU1}.)

\subsection{Non-Abelian gauge groups}
\emph{The group} $SU(2)$.--- The irreps $j$ are labelled in the set $j \in \{0,\frac{1}{2},1,\frac{3}{2},2,.\ldots \}$ and the irreps are self dual, $j^*=j$. The branching rule at vertex with two incoming edges of irreps $j_1$ and $j_2$ and one outgoing edge of irrep $j_3$ satisfies $|j_1-j_2|\leq j_3\leq j_1+j_2$, implying branching strings nets can occur on the network. One must use the full expression for the matrix elements of $\hat{B}^j_p$ where the $F$ matrices are proportional to the Wigner $6-j$ symbols:

\begin{equation}
\left(F^{a,b,c}_{d}\right)_{e,f}=(-1)^{a+b+c+d} \sqrt{(2e+1)(2f+1)}\left\{\begin{array}{ccc}a & b & e \\c & d & f\end{array}\right\}.
\end{equation}

\end{document}